ABSTRACT

Title of Dissertation:  DATA-DRIVEN RISK MODELING FOR INFRASTRUCTURE PROJECTS USING ARTIFICIAL INTELLIGENCE TECHNIQUES

Abdolmajid Erfani, Doctor of Philosophy, 2023

Dissertation directed by:  Professor Qingbin Cui, Civil & Environmental Engineering

Managing project risk is a key part of the successful implementation of any large project and is widely recognized as a best practice for public agencies to deliver infrastructures. The conventional method of identifying and evaluating project risks involves getting input from subject matter experts at risk workshops in the early phases of a project. As a project moves through its life cycle, these identified risks and their assessments evolve. Some risks are realized to become issues, some are mitigated, and some are retired as no longer important. Despite the value provided by conventional expert-based approaches, several challenges remain due to the time-consuming and expensive processes involved. Moreover, limited is known about how risks evolve from *ex-ante* to *ex-post* over time. How well does the project team identify and evaluate risks in the initial phase compared to what happens during project execution? Using historical data and artificial intelligence techniques, this study addressed these limitations by introducing a data-driven framework to identify risks automatically and to examine the quality of early risk registers and risk assessments. Risk registers from more than 70 U.S. major transportation projects form the input dataset.

Firstly, the study reports a high degree of similarity between risk registers for different projects in the entire document of the risk register, and the probability and consequence of each risk item, suggesting that it is feasible to develop a common risk register. Secondly, the developed data-driven model for identifying common risks has a recall of over 66% and an F1 score of 0.59 for new projects, i.e., knowledge and experience of similar previous projects can help identify more than 66% of risks at the start. Thirdly, approximately 65% of *ex-ante* identified risks actually occur in projects and are mitigated, while more than 35% do not occur and are retired. The categorization of risk management styles illustrates that identifying risks early on is important, but it is not sufficient to achieve successful project delivery. During project execution, a project team demonstrating positive doer behavior (by actively monitoring and identifying risks) performed better. Finally, this study proposes using a data-driven approach to unify and summarize existing risk documents to create a comprehensive risk breakdown structure (RBS). Study results suggest that acquired knowledge from previous projects helps project teams and public agencies identify risks more effectively than starting from scratch using solely expert judgments.

DATA-DRIVEN RISK MODELING FOR INFRASTRUCTURE PROJECTS
USING ARTIFICIAL INTELLIGENCE TECHNIQUES

by

Abdolmajid Erfani

Dissertation submitted to the Faculty of the Graduate School of the

University of Maryland, College Park, in partial fulfillment

of the requirements for the degree of

Doctor of Philosophy

2023

Advisory Committee:
Dr. Qingbin Cui, Chair
Dr. Vanessa Frias-Martinez, Dean's representative
Dr. Gregory Baecher
Dr. Mark Austin
Dr. Young Hoon Kwak

# Dedication

To

My beloved Maryam, your love, support, and encouragement have kept me going through the difficulties and obstacles of this journey.

Mahmoud and Fatemeh, my parents, shaped me into who I am today through their love, hard work, support, and dedication.

# Acknowledgements

My deepest gratitude goes to Professor Qingbin Cui, who expertly guided me through my graduate education and was the driving force behind all my achievements in four years of research. Thanks to his unwavering support and trust in me, I was able to complete this dissertation, and his generosity made my time at UMD enjoyable. He provided guidance and directions to follow when I was feeling stuck and unable to find a way out of my problems. It was a privilege to have him as an advisor, mentor, and most importantly friend.

A special thanks is extended to my dissertation committee: Professor Baecher, Professor Kwak, Professor Frias-Martinez, and Professor Austin. Throughout the dissertation development process, their support and guidance proved invaluable. Having the opportunity to work with highly reputed scholars on my dissertation committee enabled me to learn both professionally and personally from them. Last but not least, I am deeply grateful to my wife, parents, friends, and families. I have been encouraged by them to believe I can succeed in any endeavor I choose. Thanks for all your support along the way.

# Table of Contents

# List of Tables

# List of Figures

# List of Abbreviations

| Abbreviation | Explanation |
|---|---|
| ASCE | American Society of Civil Engineers |
| AI | Artificial Intelligence |
| BERT | Bidirectional Encoder Representations from Transformers |
| DB | Design Build |
| DBB | Design Bid Build |
| DBFM | Design Build Finance Maintenance |
| DBFOM | Design Build Finance Operate Maintenance |
| DBM | Design Build Maintenance |
| DL | Deep Learning |
| DOT | Department of Transportation |
| FHWA | Federal Highway Administration |
| FN | False Negative |
| FP | False Positive |
| FSA | Finite State Automaton |
| ISMP | Information Source for Major Transportation projects |
| ML | Machine Learning |
| NEPA | National Environmental Policy Act Review |
| NLP | Natural Language Processing |
| PPP | Public Private Partnership |
| RBS | Risk Breakdown Structure |
| RLA | Risk Life-cycle Automaton |
| ROW | Right of Way |
| SBERT | Sentence - BERT |
| SMEs | Subject Matter Experts |
| TF-IDF | Term Frequency - Inverse Document Frequency |
| TP | True Positive |
| WSDOT | Washington State Department of Transportation |

# CHAPTER 1: INTRODUCTION

## 1.1 Risk Management in Infrastructure Projects

United States' economy and citizens' quality of life are reliant on the surface transportation network. According to American Society of Civil Engineers' (ASCE) 2021 report, the national grade for infrastructures in the country is C-. In its assessment of the infrastructure over recent decades, the ASCE emphasizes the need for government investment in major transportation projects (ASCE 2021). Federal Highway Administration (FHWA) defines major projects as projects requiring more than $500 million in federal funding (FHWA 2021). Despite their keen interest in participating in major public transportation projects or updating aging infrastructures, government agencies are generally not able to finish them on time under uncertain conditions or keep their expenses within budgets. Specifically, multiple stakeholders, a broader geographical area, and long-term projects have made infrastructure projects increasingly complex (Afzal et al. 2021; Creedy et al. 2010; El-Sayegh and Mansour 2015). Some of the failed major U.S. public projects include California's high-speed rail, Maryland's Purple Line, South Carolina's I-73, and Texas's SH-288 (Linton 2018; Slowey 2019; Tuohy 2020).

Risk management is considered a best practice for public agencies for ensuring successful project delivery by simultaneously considering time, cost, quality, safety, and environmental sustainability (Abdelgawad and Fayek 2010; Cui and Erfani 2021; Erfani and Tavakolan 2020). Risk detection is crucial to the risk management process because it forms the basis of risk assessment, response, and allocation (Jung and Han 2017). It is common practice in risk management to rely heavily on Subject Matter

Experts (SMEs) for input during the identification and evaluation phases (Siraj and Fayek 2019). In major transportation projects, risk workshops are routinely conducted to develop risk registers that document all risks identified (Leva et al. 2017).

Based on the FHWA risk management guideline, risk management processes in highway projects include five main stages: identification, assessment, planning, allocation, and monitoring (Molenaar 2006). Risk identification is the mechanism of determining which risk items may affect the project. Brainstorming, checklist analysis, literature and documentation review, workshops, Delphi technique, questionnaire survey, root cause analysis, and cause and effect diagrams are expert judgment-based tools to detect risks in highway construction projects (Taroun, 2014). Risk assessment centers on measuring the importance of each risk. Identified risks are compared in different scales of probability, cost, and schedule impact to find those risks that need more attention (Islam et al. 2019; Heravi et al. 2021). Risk assessment is conducted in both quantitative and qualitative approaches. Qualitative analysis involves assessing the probability and impact using risk matrix and categorical scales such as High, Medium, and Low (Siraj and Fayek 2019; Molenaar 2006). The quantitative analysis for probability and impact is conducted using numerical scales. Then risk allocation aims to determine which project party is most appropriate to bear any risk. The contract is used as the procedure to allocate the risks and consider the ex-post remedies for the risks if they occur (Nguyen et al. 2018; Wang et al. 2022). Finally, monitoring and controlling risks are repeated during the project implementation to update all the available information.

Many scholars have studied the risk management process in infrastructure projects, mostly focusing on the ex ante perspective and solely rely on expert judgment (Hastak and Shaked 2000; Iqbal et al. 2015). All these steps are completed before the project has started or when the project has not progressed significantly. While these studies focus on ex ante risks, few have focused on retrospective analysis of risks remaining after the project is completed, that is, ex post (Figure 1).

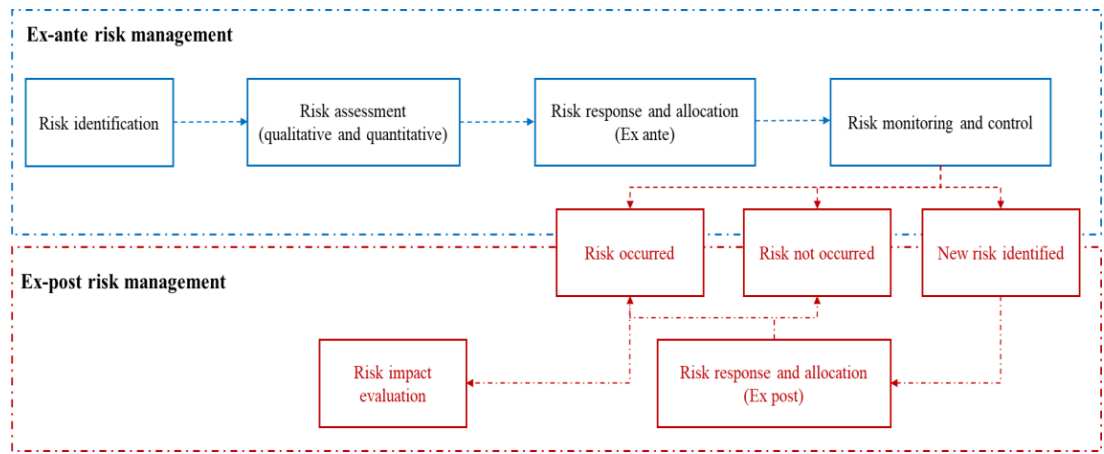

Figure 1. Processes of ex ante and ex post risk management

## 1.2 Problem Statement and Research Need

While the expert judgement-based approach dominates construction, the application process and its costs are typically lengthy and expensive (Gondia et al. 2020; Sanni-Anibire et al. 2020; Somi et al. 2021). Also, there are cognitive and subjective judgment biases associated with developed risk registries for public agencies with limited experience with major transportation projects (Duijm et al. 2015; Montibeller et al. 2015). Additionally, an expert group with different backgrounds and knowledge may have different perspectives on risk assessment, which makes achieving a decision more difficult (Erfani et al. 2021a; Mohammadi et al. 2022; Monzer et al. 2019). These challenges must therefore be addressed in current risk management practices.

As well, most of the literature focuses on *ex-ante* risk identification and assessment performed at the beginning of the project, but there are gaps in measuring the performance of risk identification and its impact on project delivery. When compared with project execution, how well does the project team identify and evaluate risks in the initial phase? At the end of the project, what proportion of the risks identified *ex ante* have been realized, and what portion has been dismissed? Risk registers can be more effective if we develop a better understanding of this dynamic.

**1.3 Research Objectives**

As a result of discussed problems, this study proposes evaluating the uniqueness of risk registers through a data-driven approach. Next, it develops a predictive risk detection model based on historical data extracted from previous projects with similar features to create initial risk templates, based on risk similarity among risk registers. Furthermore, using automata theory, the study proposes a framework for measuring risk management performance. The risk life cycle in construction projects follows a state-transition logic that can be formalized mathematically as a finite state automaton. The automaton defines risk status states and transitions among those states as a project progresses. The study introduces new terms for risk management styles based on the project team performance during the initial and project execution stages. Finally, this study introduces a comprehensive data-driven risk breakdown structure with demonstrating the importance of evaluating the relationship between risks on a network basis. Figure 2 shows the dissertation framework.

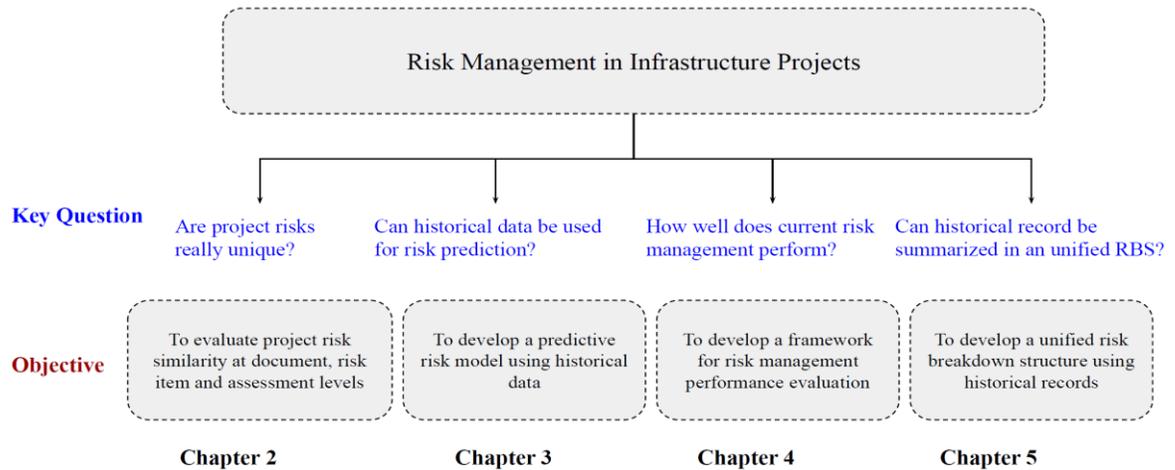

Figure 2. Dissertation research objectives

In summary, this study aims to accomplish the following objectives:

1. Evaluating the uniqueness of risk registers in infrastructure projects by calculating similarity indexes to answer the question "RQ1: Are project risks really unique?"

2. Developing a predictive data-driven risk detection model through the implementation of historical data extracted from previous projects with similar features to create initial risk templates to answer the question "RQ2: Can historical data be used to predict project risks?"

3. Developing a framework to evaluate project risk performance by comparing risk register documents before, during, and after project execution to answer the question "RQ3: How well does current risk management perform?"

4. Categorizing project team risk management style and behaviour to answer the question "RQ4: Does risk management style affect project delivery performance?"

5. Developing a unified risk breakdown structure to answer the question "RQ5: Can historical risk report can be used to develop a unified risk breakdown structure as a preliminary risk identification framework?"

**1.4 Natural Language Processing and Deep Learning**

Artificial Intelligence (AI) techniques, in particular, Natural Language Processing (NLP) models, formed the foundation for this study. Most risk registers include a huge amount of text and a multitude of risk items, and the project team describes the risks in its own words. Hence, manual comparison requires extensive time and resources, which can be efficiently accomplished by applying advancements from NLP techniques.

NLP techniques first transfer human language to the structured text and then to numeral data for further analysis and modeling. Recent studies on NLP techniques significantly improved efficiency by using new algorithms that focus on semantic meaning and context (Di Giuda et al. 2020; Erfani and Cui 2021; Erfani et al. 2023a). Similarly, modern Machine Learning (ML) and Deep Learning (DL) algorithms effectively provide efficient methods to convert unstructured text data to machine-readable formats for analyses (Lauriola et al. 2022). With the emergence of word embedding models (neural networks) that convert higher dimension mathematical spaces into shallow word embodied vectors, tremendous opportunities have opened up (Zhang 2019; Zhong et al. 2020). While capturing the semantic meaning behind words, models show each specific word in the corpus. Major artificial intelligence companies such as Google and Facebook established large pre-trained word embedding vectors. Google's research team introduced Word2Vec as the first pre-trained vector model (Mikolov et al. 2013). Facebook then released a pre-trained model, FastText (Bojanowski et al. 2017), and the Stanford NLP group presented the GloVe model (Pennington et al. 2014). Word embedding models are an application of DL, converting text into a vector by considering the semantic meaning behind the word. Several pre-trained word embedding models contain a set of two-layer neural networks to generate vectors for each word, for example, Word2Vec, one of the popular and powerful DL models, was trained using millions of words from Google news and, each word was converted into a 300-dimension vector based on the meaning.

**1.5 Dissertation Outline**

Figure 3 illustrates the dissertation's structure. The dissertation is organized as follows: Chapter 2 investigates the uniqueness of risk registers among infrastructure projects. Results will demonstrate that data-driven approaches can facilitate the development of a common risk register while still allowing project teams to focus on their unique risks. A systematic comparative analysis based on NLP and a state-of-the-art deep learning algorithm named Word2vec is used to calculate the similarity index at three levels, i.e., the entire document of the risk register, individual risk items, and probability and consequence of each risk.

In chapter 3, the study presents a predictive risk modeling based on historical data instead of solely depending on expert judgment. A data-driven approach utilizes NLP and word embedding models to detect risks with similar terminologies from past risk registers. By considering both the prevalence and cost/time implications of risks and specific project characteristics, the model is able to capture critical risks. Furthermore, the model has been tested with five projects regarding risk prediction.

In chapter 4, performance metrics and framework for evaluating the project team's risk identification performance are presented. The framework is informed by automata theory to define risk states and transition functions to track risk life-cycles. Metrics are designed to determine the percentage of risks that occurred and were dismissed during the course of the project. The metrics are classified into three groups based on the performance of risk identification in the total, initial, and project execution phases. Finally, the study introduced new terms to categorize a project teams' risk style based on their planning and doing behaviors. The fifth chapter covers developing a common risk breakdown structure as a preliminary risk identification framework and evaluating risk interdependencies based on co-occurrence of risk items in historical risk data. A summary of key findings and suggestions for future research were presented in Chapter 6, along with managerial implications.

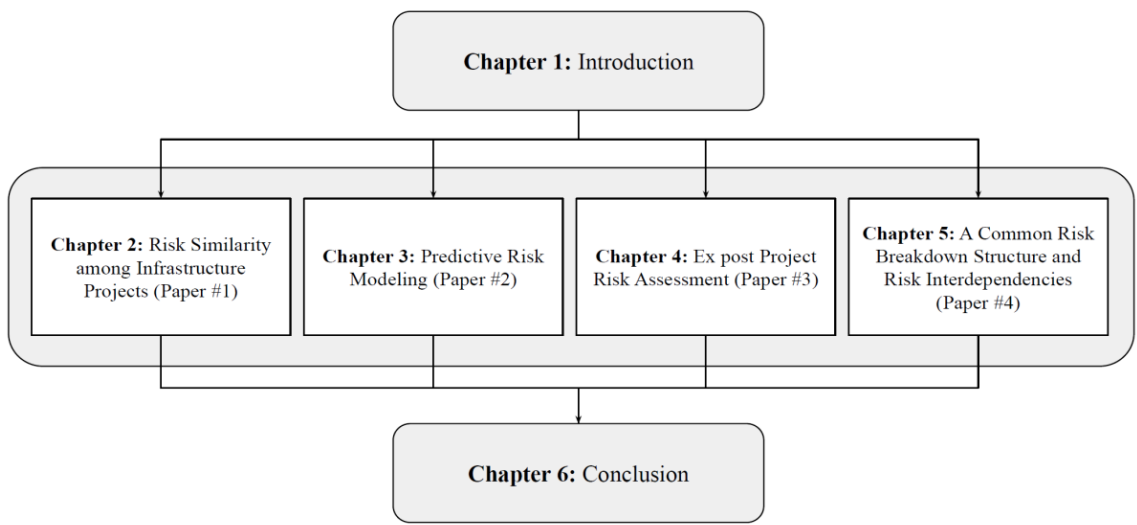

Figure 3. Dissertation structure

# CHAPTER 2: RISK SIMILARITY AMONG INFRASTRUCTURE PROJECTS

The contents of Chapter 2 are published in the Journal of Construction Engineering and Management, ASCE.

**Citation**: Erfani, A., Cui, Q., & Cavanaugh, I. (2021b). An Empirical Analysis of Risk Similarity among Major Transportation Projects Using Natural Language Processing. *Journal of Construction Engineering and Management*, *147*(12), 04021175.

## 2.1 Abstract

Risk management is widely recognized as a best practice for public agencies to ensure the successful implementation of major transportation projects. The conventional approach to identify and evaluate project risks is dominated by getting input from subject matter experts at risk workshops. However, the uniqueness of such a risk assessment approach remains unexamined. How different are the risks among various projects? Does the risk register reflect the unique feature of a project? The goal of this study is to measure the similarity of project risks across various groups by evaluating 70 major transportation projects delivered under various methods. The similarity index is calculated at three levels, i.e., the entire document of the risk register, individual risk item, and the probability and consequence of each risk using a systematic comparative analysis based on NLP and a state-of-the-art deep learning algorithm named Word2vec. The study reports a high similarity of risk registers among different projects at all three levels. The analysis does show a lower similarity of risk registers for PPP projects. The primary contributions of this study are (1) develop a new approach to analyze the risk registers at the project level as the main output of risk management practice. (2) establish the relation of risk uniqueness and project delivery method in transportation projects. Results suggest that a data-driven approach may be possible to help project teams develop a common risk register while allowing the teams to focus on each project's unique risks.

**2.2 Research Design and Data**

As each project is unique; project risk should reflect this unique nature. To evaluate the uniqueness of risk management, this study utilizes NLP algorithms to calculate the similarity of risk register documents in three different levels. Main purpose compares differences and similarities between disparate projects' risk registers. Risk registers document all the SME's workshop output. Similarities provide a novel solution to reduce project teams' efforts to conduct the risk analysis process in future steps.

Three Level of Similarity Comparison are as follows:

- Document Level

The first similarity comparison is conducted at the document level. The purpose is to consider all the data inside a risk register document including risk categories, risk names, and risk descriptions. In this level of comparison, the words are considered without capturing the meaning and sequence. The result indicates the similarity in the context of risk register documents.

- Individual Risk Item Level

Second phase assesses individual risk items, incorporating semantic meaning behind each word. Sometimes project teams use different language to identify the same risk item. The purpose is to match those similar risks with different terminology in various projects. The result illustrates how similar the identified risks in different risk registers are overall.

- Evaluation Level

The third layer of comparison is used to compare the risk assessment for those matched risks in the second level of comparison. The project team evaluates the probability, cost impact, and schedule impact of identified risks in each project. In this step, these evaluations will be compared in both quantitative and qualitative risk analysis. The result indicates how project teams evaluate the same risks in different projects similarly in terms of consequences.

2.2.1 Data Collection and Preprocessing

A dataset of risk registers of major infrastructure projects served as the primary source data of this study. The dataset comprises 70 major transportation projects with different project delivery methods and contract values from various states of the United States that were mostly delivered in the last decade. Table 1 lists detailed information about selected projects. These projects include highway projects which divide into two groups of (A) traditional delivery methods group (DB, DBB) and (B) PPP projects. In group A the contract award date reflects the first major awarded contract and in group B it indicates the major concessioner contract award date. Also, Figure 4 outlines the process of preparing the dataset. The dataset includes risk categories, risk names, risk descriptions, and risk evaluation. Each risk evaluation includes probability, cost impact, and schedule impact. In order to compare the risk evaluation, the data is converted in a standard scale based on Virginia risk management guidelines (Partnerships 2015). The probability, normalized cost impact, and schedule impact converted to 1-5 standard Likert scales for more efficient quantitative comparison rather than using continuous numbers. In the same way by using (Partnerships 2015) scales, the combined probability, cost impact, and schedule impact are converted to the three levels of quantitative assessment including high, medium, and low assessments. This standardization of items prepares the dataset for further similarity calculations.

Table 1. Projects Detailed Information

| Project ID | Jurisdiction | Delivery Method | Contract value (Million $) | Number of risks | Contract award year |
|---|---|---|---|---|---|
| Group A: DB/DBB projects | | | | | |
| A-1 | AZ | DB | 1773 | 54 | 2016 |
| A-2 | DC | DB | 669 | 28 | 2017 |
| A-3 | FL | DB | 1024 | 105 | 2007 |
| A-4 | FL | DB | 1004 | 67 | 2015 |
| A-5 | NY | DBB | 1079 | 55 | 2012 |
| A-6 | FL | DB | 509 | 126 | 2016 |
| A-7 | WI | DBB | 1625 | 135 | 2009 |
| A-8 | AL | DBB | 746 | 18 | 2013 |
| A-9 | CA | DBB | 166 | 69 | 2016 |
| A-10 | CA | DB | 863 | 22 | 2014 |
| A-11 | CA | DBB | 301 | 19 | 2008 |
| A-12 | CO | DBB | 610 | 20 | 2018 |
| A-13 | FL | DB | 852 | 233 | 2016 |
| A-14 | IL | DB | 906 | 18 | 2013 |
| A-15 | IA | DBB | 1131 | 17 | 2017 |
| A-16 | KY | DBB | 583 | 35 | 2014 |
| A-17 | MD | DBB | 814 | 44 | 2016 |
| A-18 | MI | DBB | 2950 | 34 | 2015 |
| A-19 | MN | DBB | 647 | 71 | 2013 |
| A-20 | MS | DBB | 610 | 38 | 2012 |
| A-21 | NV | DBB | 1237 | 113 | 2017 |
| A-22 | NY | DBB | 1079 | 110 | 2014 |
| A-23 | NY | DBB | 953 | 13 | 2010 |
| A-24 | NY | DB | 4825 | 11 | 2012 |
| A-25 | NC | DB | 731 | 104 | 2011 |
| A-26 | CA | DBB | 850 | 32 | 2014 |
| A-27 | PA | DBB | 678 | 28 | 2015 |
| A-28 | PA | DBB | 1641 | 17 | 2014 |
| A-29 | TX | DB | 1585 | 36 | 2011 |
| A-30 | TX | DB | 693 | 38 | 2017 |
| A-31 | VA | DB | 921 | 86 | 2016 |
| A-32 | WA | DB | 534 | 243 | 2014 |
| A-33 | WI | DBB | 1202 | 35 | 2012 |
| A-34 | WI | DBB | 410 | 17 | 2014 |
| A-35 | WI | DBB | 1550 | 15 | 2012 |
| A-36 | CA | DBB | 584 | 38 | 2007 |
| A-37 | WA | DB | 563 | 123 | 2010 |
| A-38 | TX | DB | 743 | 22 | 2015 |
| A-39 | CA | DBB | 986 | 134 | 2003 |
| A-40 | IA | DBB | 2629 | 49 | 2008 |
| A-41 | IL | DBB | 3628 | 11 | 2018 |
| A-42 | TX | DBB | 4922 | 245 | 2011 |
| A-43 | MS | DBB | 1296 | 81 | 2009 |
| A-44 | CA | DBB | 1792 | 140 | 2010 |

**Table 1. (Continued)** Projects Detailed Information

| Project ID | Jurisdiction | Delivery Method | Contract value (Million $) | Number of risks | Contract award year |
|---|---|---|---|---|---|
| Group A: DB/DBB projects | | | | | |
| A-45 | DE | DBB | 860 | 33 | 2015 |
| A-46 | CA | DBB | 817 | 10 | 2013 |
| A-47 | IL | DBB | 745 | 40 | 2014 |
| A-48 | CA | DB | 1421 | 32 | 2013 |
| A-49 | NV | DB | 955 | 51 | 2015 |
| A-50 | CA | DB | 1492 | 27 | 2012 |
| A-51 | CA | DB | 1910 | 116 | 2016 |
| Group B: P3 Projects | | | | | |
| B-1 | OH | DBFOM | 3564 | 36 | 2016 |
| B-2 | CO | DBFOM | 1204 | 101 | 2017 |
| B-3 | FL | DBFOM | 4854 | 107 | 2014 |
| B-4 | IN | DBFOM | 1064 | 73 | 2012 |
| B-5 | MI | DBFM | 1137 | 97 | 2018 |
| B-6 | NY & NJ | DBFM | 1116 | 131 | 2013 |
| B-7 | TX | DBFOM | 899 | 63 | 2015 |
| B-8 | CA | DBFOM | 488 | 42 | 2010 |
| B-9 | FL | DBF | 880 | 181 | 2018 |
| B-10 | GA | DBF | 764 | 29 | 2015 |
| B-11 | GA | DBF | 834 | 19 | 2013 |
| B-12 | PA | DBFM | 1213 | 18 | 2014 |
| B-13 | VA | DBFOM | 3863 | 19 | 2016 |
| B-14 | FL | DBFOM | 974 | 43 | 2008 |
| B-15 | VA | DBFOM | 1400 | 174 | 2006 |
| B-16 | TX | DBM | 1066 | 29 | 2015 |
| B-17 | NC | DBFOM | 661 | 107 | 2004 |
| B-18 | TX | DBFOM | 2146 | 56 | 2009 |
| B-19 | TX | DBFOM | 1793 | 55 | 2009 |

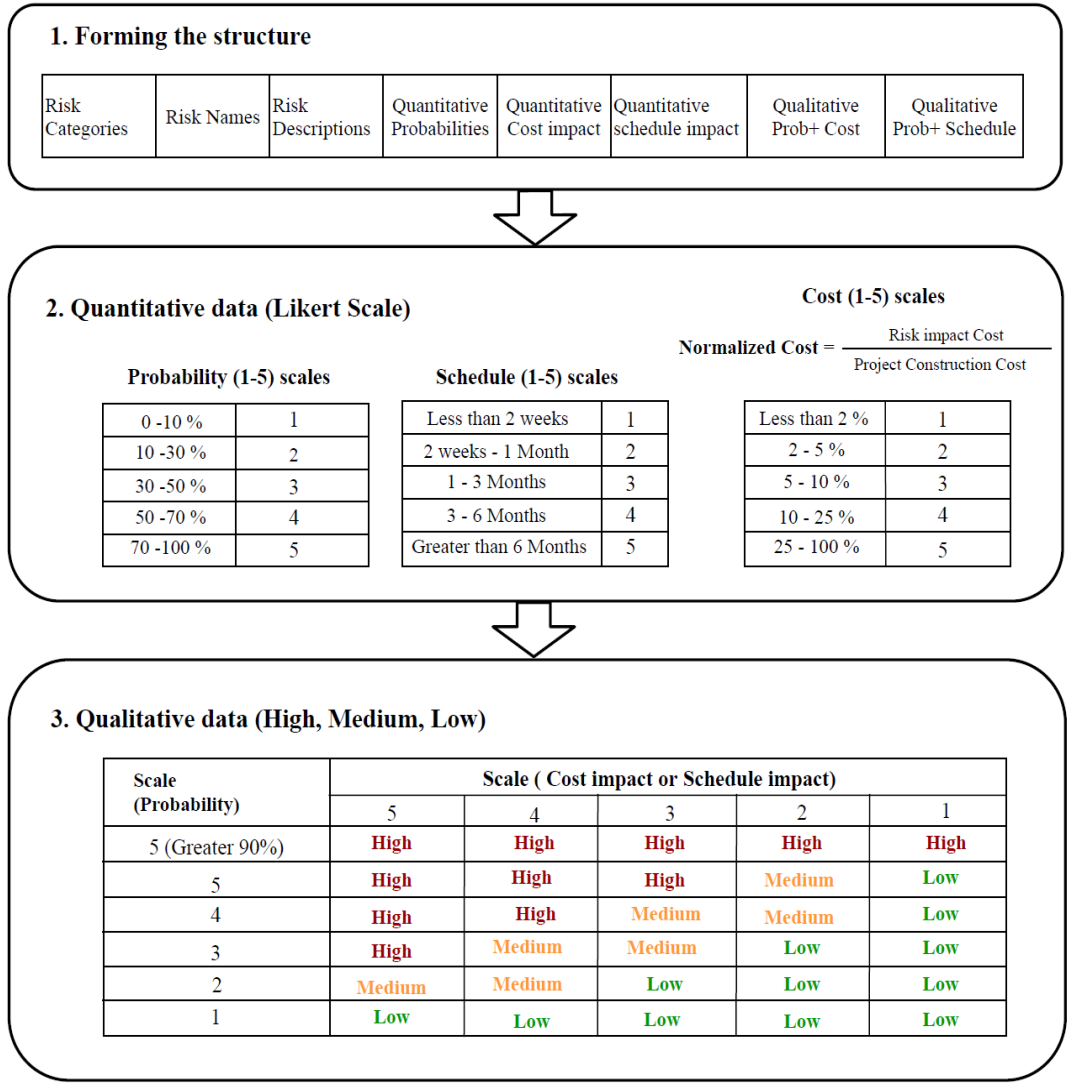

Figure 4. Dataset preparation

2.2.2 Similarity Calculation

Automatic similarity calculation between two documents is one of the key NLP applications. The most common approach is to convert text to numeral numbers based on their features to compute the level of similarity in a vector space (Shahmirzadi et al. 2019). There are two common approaches for representing a text document in a vector space which was used in this study.

**Bag of Words**

Bag of words offers the simplest method of representing text documents in numeral vector space (Hassan and Le 2020). This method does not consider the actual meaning behind the words and their orders in the sentences. All the words in the corpus are considered as one element and based on whether the corpus includes that word or not, represents it in a vector space. Another revised format is using the frequency of the words in the whole corpus and then converting the sentence to the vector space. Term frequency- invert document frequency (TF-IDF) is one of the most basic text vectorizations in this group (Shahmirzadi et al. 2019). The logic behind TF-IDF is to decrease the importance of common words which are repeated a lot in a document and cannot help to distinguish the difference between two documents (Sidorov 2019). TF-IDF score for each term inside the documents is calculated as follows:

$$TF - IDF_{score} = \frac{n_t}{N} \times (1 + \log \frac{k}{k_t})$$

(1)

Where:

$n_t$: Number of occurrence of terms t in the document

$N$: Total number of terms in the document

$k$: Total number of documents

$k_t$: Number of documents containing the term t

This approach is used in the first level of similarity comparison in this study. All the terms inside a risk register including risk categories, names, and descriptions are used to calculate the TF-IDF scores. After removing the stop words, the vector which represents the whole risk register document is calculated. Then, the degree of similarity between two documents is computed using the cosine similarity term. Cosine similarity is a robust metric to calculate the level of similarity by measuring the cosine of the angle between two vectors as follows (Fan and Li 2013):

$$Similarity\ (Doc1,\ Doc2) = Cosine\ (v, w) = \frac{V.W}{||V||||W||} = \frac{\sum_{i=1}^{k} V_i.W_i}{\sqrt{\sum_{i=1}^{k} V_i^2} \times \sqrt{\sum_{i=1}^{k} W_i^2}} \qquad (2)$$

Where:

V: vector representing the first document

W: vector representing the second document

Pairwise comparison of risk registers in document level describes contextual similarity. Similarity level shows the overall matching of risk registers considered as one document.

**Word Embedding**

Word embedding models utilize neural networks and artificial intelligence to generate a vector to represent the semantic meaning behind every unique word inside the corpus (Sidorov 2019). Several pre-trained word embedding models exist, including Word2vec (provided by Google) (Mikolov et al. 2013), GloVe (provided by Stanford NLP group) (Pennington et al. 2014). The most important advantage of the Word2vec model is to capture the syntactic and semantic word relationships and meanings (Kim and Chi 2019). Because of the small size of words in the corpus, Word2vec, a well-known pre-trained Word embedding model is employed in this study. This approach is used in the second layer of similarity comparison. Each risk item matched to the highest similar risk item from the second risk register document using Word2vec vectorization and cosine similarity calculation. The process includes lowercasing, removing stop words, tokenizing the risk items, converting each word to a vector using Word2vec, calculating the average vector representing the risk item, and finally computing the cosine similarity. For example, risk item 1 is selected from the first risk register and then compared to all risk items in the second risk register. Then, risk item 2 is selected from the second risk register as the best match to explain one example of similarity calculation. Figure 5 describes the similarity calculation at the second level. After the completion of the matching process, the average similarity indexes indicate the overall similarity of risk registers in the risk level between two documents.

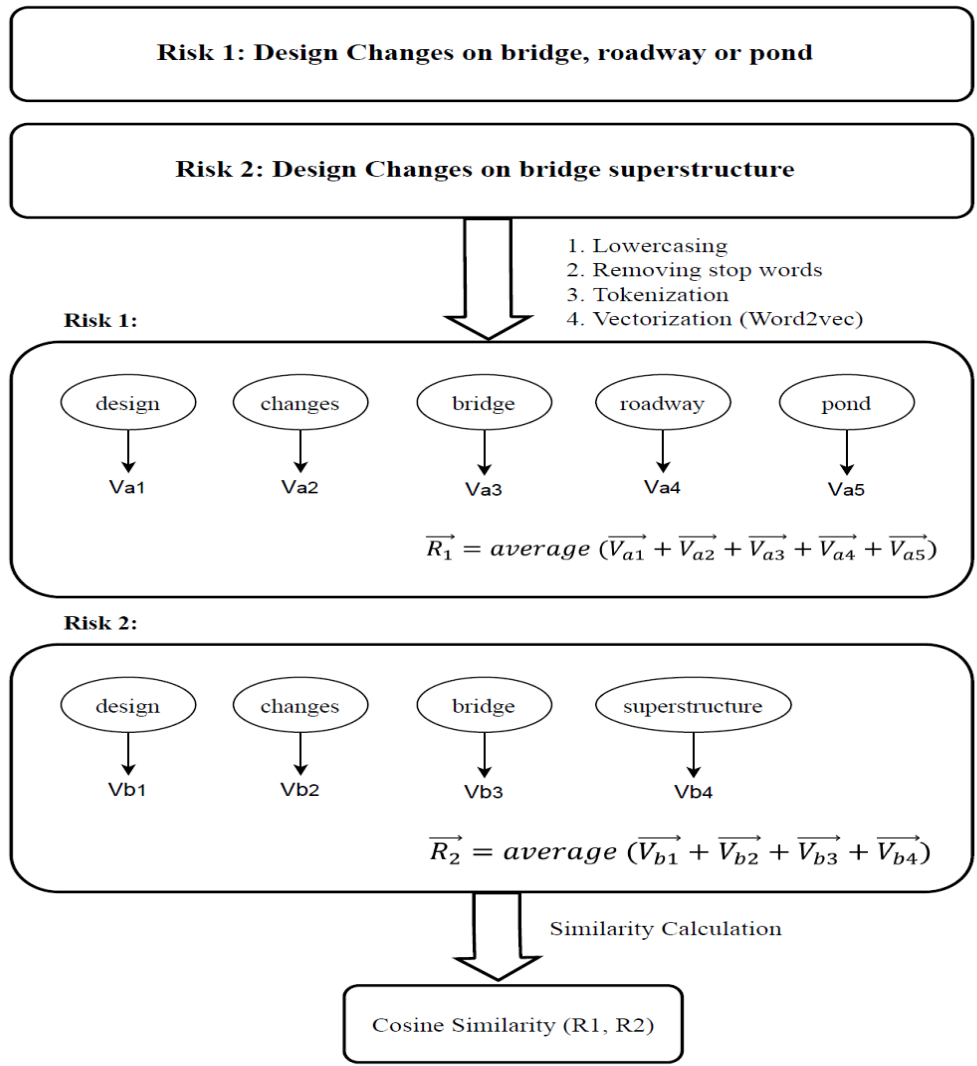

Figure 5. Cosine similarity calculation example using Word2vec

The next step of similarity calculation, pooling approach, considers one risk register on one side and compares it to all other risk registers on the other. The result of this comparison will indicate what the possibility of finding at least a similar risk item in the individual risk level is from a large pool of individual risk items. The emphasis here focuses more on individual risk items rather than the entire project.

Output of the prior matching process determines the next level of similarity comparison. Firstly, the probability, cost impact, schedule impact is compared separately in the quantitative Likert scale (1-5) based on the following equation using the distance similarity index which converts to the percentage of similarity. The same process is completed for probability, cost impact, and schedule impact comparison.

$$Similarity = 1 - Distance\ index = [1 - \left(\frac{|x_1 - x_2|}{4}\right)] * 100 \qquad (3)$$

Where: $x_1$ is risk evaluation (probability or cost impact or schedule impact) of risk item 1 and $x_2$: risk evaluation (probability or cost impact or schedule impact) of risk item 2

Secondly, the combined probability, cost impact, combined probability, and schedule impact are compared for matched risks on a qualitative scale. The percentage of matching in the high, medium, and low scales is reported. In this comparison, when both evaluations are similar, the similarity will be considered as 100% if not 0%.

In summary, the risk register documents of major infrastructure projects are compared based on three levels of similarity. First, at the document level using the TF-IDF method, second, at the risk level using the Word2vec method, and finally, at the evaluation level using distance and matching similarity indexes. The result will indicate the level of uniqueness in risk register documents.

**2.3 Results and Discussion**

2.3.1 Similarity at Document Level

Pairwise comparisons of risk registers at the document level describe that they are highly similar in the context. Without capturing the meaning and order of words inside each document, the average of cosine similarity at this level is 0.67 for the group A including traditional delivery method projects. The results are in the range of 0.39 to 0.96, computed using TF-IDF methods as explained in bag of words section. For PPP projects, the average similarity is 0.52 and the range is 0.31 to 0.98.

The comparison at this level shows that when selecting two random risk registers from the project delivered under DBB or DB, an average of 67% similarity context will be observed. However, for PPP projects, risk registers show high uniqueness, causing this similarity to be reduced to 52%. At the end of each section, a simple standardized T-test is conducted to compare the mean difference in P3 and DB/DBB groups. At the document level, the P-value is 2.36e-11 (less than 0.05) which indicates there is a significant difference (De Winter 2013). This difference among PPP and DB/DBB groups indicates that more project-specific words and phrases are typically used in risk documents for the PPP projects group.

2.3.2 Similarity at Individual Risk Item Level

The similarity calculation at the document level shows that risk register documents use similar words. The second level of comparison is conducted at an individual risk item level to show how similar the identified risks are. The pre-trained Word2vec model is used to match the risks based on the meaning behind words and calculation of cosine similarity. Sometimes, project teams use different terminology for similar risk items.

Therefore, the purpose of similarity comparison at this level automatically locates matched risks.

Firstly, the average of cosine similarity matching among those risk items considering the risk names is 0.51 for DB/DBB delivery project group. By considering the risk names plus risk description as one element for each risk item, the average is 0.64. This result illustrates that project teams use more similar words to propose the risk descriptions in risk register documents. However, based on the authors' evaluation of a sample of matching results, the intelligence of matching using the risk names alone is greater than risk name besides risk description. The first reason behind this is that project teams do not follow the same structures for describing risk descriptions, and the second reason is that when the number of the words increases, the accuracy of matching decreases. Therefore, we decided to make our model and analysis based on the risk names. The result for PPP projects is 0.48 similarity on average. Like the previous section, the PPP project risk registers are more unique in the individual risk item level as well as the P-value of the T-test is 0.00123 which supports the significant difference between two groups. The detailed result of comparison for randomly selected 7 projects in each group at this level is presented as a heatmap in Figure 3. In Figure 6, in each comparison, the risk items in the horizontal project are used to match to risks in the vertical project. Since projects contain an unequal number of risk items, the average comparison of A to B may not return results equal as B to A.

Secondly, based on the result, a perfect matching of when the similarity index is 1 occurred, meaning that the same risk item was repeated exactly in different projects. For those comparisons that have more than 0.7 similarity index, a strong matching is observed. Also, based on the result of when the cosine similarity is more than 0.5, there is a meaningful connection between two risk items.

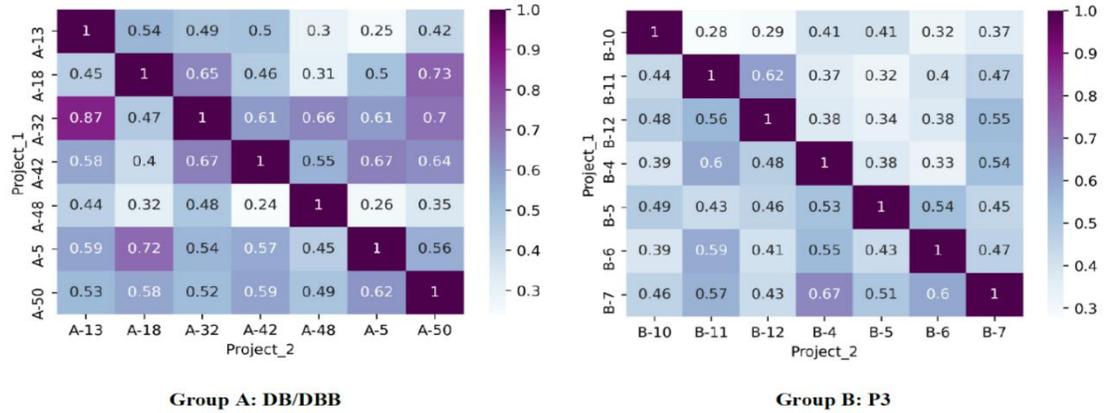

Figure 6. Result of project pairwise similarity comparison at risk level

Table 2 provides some examples for risk items in each level of similarity. Note that each risk item belongs to a different project and this table is a simple example of matching results.

Table 2. Risk Matching Examples

| Risk 1 | Similarity | Risk 2 |
|---|---|---|
| Delay in ROW document internal approval process | 1 | Delay in ROW document internal approval process |
| Contractor delays and default | 1 | Contractor delays and default |
| Encountering unexpected subsurface conditions | 1 | Encountering unexpected subsurface conditions |
| Utility relocation may not happen in time | 0.965 | Utility relocation may not happen on time |
| Determination of secondary impacts to wetlands | 0.963 | Determination of wetlands impacts |
| Changing geotechnical conditions | 0.865 | Changing geotechnical conditions in bridge approaches |
| Negative community impacts cause delays | 0.832 | Negative community impacts expected |
| Unsuitable / contaminated materials | 0.816 | Unanticipated hazardous materials or contaminated soils |
| Unstable subsurface conditions | 0.784 | Encountering unexpected subsurface conditions |
| Design changes on bridge, roadway, or pond | 0.756 | Design changes on bridge superstructure |
| Delay in the ROW acquisition along | 0.738 | ROW acquisition delays construction |
| Change in contract packaging | 0.689 | Construction contract packaging |
| Concrete delivery | 0.586 | Materials delivery constraints (On-site) |
| Opportunity to get low bids due to market conditions | 0.578 | Market conditions |

Pooling approach, as mentioned, centers on individual risk items rather than entire projects. Each project is collectively compared to all other projects in that group. Then, the best match risk among those projects in the same group is selected for each risk item. The average result based on risk items similarity is shown in Figure 7 which includes detailed information about the percentage of risk matching in each level of similarity. There is a cumulative line that explains the cumulative result of pairwise similarity among each project and all other risk items together. As Figure 7 shows, the average possibility of finding at least one similar risk item (with more than 50% cosine similarity) for a new project from other projects in DB and DBB groups is 99% and 97% in PPP groups. This result proves the development of new project risk registers can borrow from previous efforts.

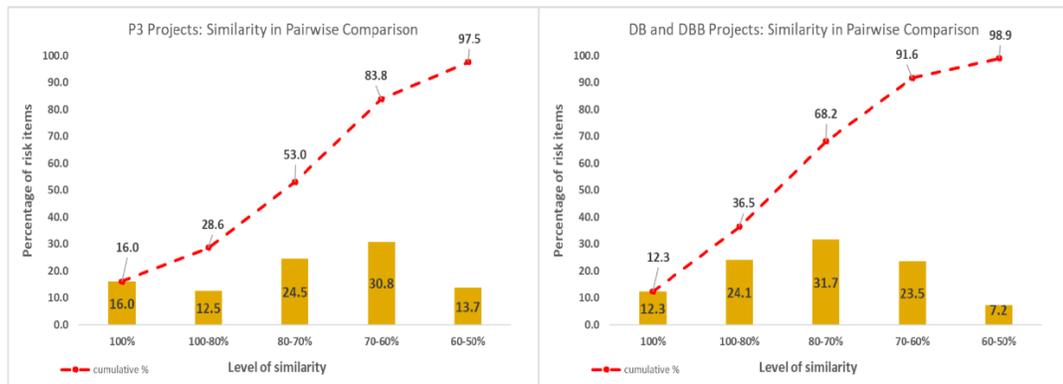

Figure 7. Result of pairwise similarity comparison at risk level using pooling approach

2.3.3 Similarity at Evaluation Level

As discussed, comparing similarity at the evaluation level identifies whether or not similar risk items in different major transportation projects follow the same quantitative and qualitative assessment in terms of probability, cost, and schedule impact. Among pairwise comparisons, risks matched with at least 0.5 cosine similarity are filtered and the distance similarity index is calculated to compare the quantitative data for risk evaluation. The similarity index is a number between 0 and 100 with a number close to 100 indicating that the risks follow the same evaluation. Table 3 provides the result of the similarity index in various levels of cosine similarity in both delivery method groups.

Table 3. Similarity at Evaluation Level

| Cosine similarity level | Probability | Cost | Schedule | Probability + Cost | Probability +Schedule |
|---|---|---|---|---|---|
| Group A: DBB/DB Projects | | | | | |
| At least 0.5 | 59.7% | 92.3% | 76.1% | 50% | 51% |
| At least 0.7 | 61.9% | 91.6% | 88.8% | 63% | 60% |
| At least 0.8 | 62.8% | 92.4% | 88.2% | 67% | 64% |
| Group B: P3 Projects | | | | | |
| At least 0.5 | 65.3% | 98.3% | 79.0% | 79% | 55% |
| At least 0.7 | 67.4% | 98.2% | 79.1% | 78% | 56% |
| At least 0.8 | 69.7% | 99.3% | 77.5% | 72% | 52% |

Results show that there is a considerable similarity at the quantitative evaluation level. All the similarity indexes exceed 50%, supporting that similar risk items among those major transportation projects return similar assessments too. Cost assessment offers greater similarity than probability and schedule analysis. Difficulty to determine cost impacts, the role of existing schedule analysis tools, and higher knowledge of the project team provide potential key reasons behind this difference. In other words, probability and schedule analysis are more project-specific than cost evaluation.

Additionally, increasing the cosine similarity will increase the similarity index. This supports that evaluations converge with higher cosine similarity. Even for cost analysis, the average similarity index closes to 100%, meaning the same cost evaluation applied for most of the matched risks with at least 0.7 cosine similarity. Finally, compared to non-PPP endeavors, average PPP projects lists include risk evaluation more unique assessment. However, level 3 results fail to support differences in levels 1 and 2. P-value based on considering probability, cost, and schedule impact of each group together calculates at 0.941876, greater than 0.05 threshold.

Similarity calculation for qualitative data also shows that the combined analysis of cost and probability in terms of high, medium, low returns similar values. Therefore, both cost and schedule analysis return similar results with schedule analysis more project specific than cost. Table 3 depicts the result of the matching index for the qualitative analysis of combined cost and schedule probabilities in different cosine similarity levels. In this level, the uniqueness of evaluation in both PPP and traditional delivery method groups is very close.

In summary, the similarity comparison in the third level indicates that both quantitative and qualitative analyses are very close. On the other hand, cost estimation of risk impacts seems more difficult for project teams rather than delay prediction. Therefore, the similarity in cost impact is more than schedule analysis. Uniqueness for both groups of delivery methods is similar and high in this level.

2.3.4 Result Validation

Cosine similarity, calculated using the Word2vec deep learning model, returns a number between 0 and 1 (0-100%). Decreasing the cosine similarity index, the associated probability mismatching risks increases. Determining meaningful threshold values drives the validity of research findings. Two members of the research team manually validated selected matched risks and assessed the model's accuracy. Each member cross-checked and upon receiving the same label from both members, items were considered as final. For discrepancies, all research team members coordinated to finalize the label. Table 4 provides the result of model validation based on a sample of 250 risk items. Accuracy calculates based on the number of accurate matches divided by all randomly selected matching risks. Results with at least 50% cosine similarity show that there is a meaningful relationship between matched risk items which supports that our model intelligently captured similar risks with different terminologies.

Table 4. Model Validation

| Cosine similarity level | Sample size | Accuracy |
|---|---|---|
| 100 | 50 | 100% |
| 80-100 % | 50 | 100% |
| 70-80 % | 50 | 86% |
| 60-70 % | 50 | 80% |
| 50-60 % | 50 | 60% |

2.3.5 Managerial Implication

Risk management is one of the key components of project management. The current practice of risk management in highway construction relies on experts' opinions and discussions. While data-driven studies to extract valuable insights from available historical data of major transportation projects have steadily increased (Erfani et al. 2021c; Hickey et al. 2022; Mohammadi et al. 2023; Morteza et al. 2023; Panahi et al. 2022), comparable research in the risk management domain remains at an early stage. However, the explosiveness of available objective and factual data creates a great opportunity to capitalize on the data for better project performance. This study introduces a new approach to understanding risks in major transportation projects using natural language processing. A large number of major transport projects have been constructed; however, there have not been any studies evaluating the uniqueness of risk management practices in these projects. Similarity calculation offers practical benefits in many ways. Creating a data-driven tool capable of generating initial risk registers drafts for future major projects, reduces effort and time. The similarity result shows that although the project is unique, the risk items are not completely unique and can easily be borrowed from similar previous projects. Therefore, considering the importance of project feature qualifiers such as delivery method, size, location, timeline, etc. to find the most similar previous projects and using NLP techniques, future project teams benefit from a proven, reliable source of risk identification and assessment.

**2.4 Conclusion**

Major transport projects contain a high level of risk and uncertainties due to the inherent characteristics of these projects. Therefore, cost overrun, and schedule delay present main problems in transportation agency project implementation. Risk management practices try to detect these challenges, evaluate them, and propose appropriate responses to manage the projects effectively. While numerous studies in the literature provide different tools and techniques to complete the risk management process, industry practices remain experience-based and rely on opinions from subject matter experts. This study develops a data-driven approach using NLP and deep learning to measure the similarity of project risks across various groups. Two groups of traditional delivery method projects (DB and DBB) and PPP comprised the main source of risk management data in this study. Analysis evaluates components from 70 major transport projects at three different levels. Results of the similarity index at the document level illustrates that more than 60% of words and context in risk registers reflect high similarity. Further, risk and evaluation levels suggest that more than 50% of risk items in the projects appear with the same qualitative and quantitative assessment. Finally, the pooling approach suggests that more than 97% of risk items in each risk register can be found in other similar projects. Cost impact estimation returns the highest similarity between cost impact analysis, schedule impact analysis, and probability analysis of the risk items with DBB/DB consistently higher values than PPP projects. Study calculations indicate that although the projects are unique, the risk register documents contain repeated items. Therefore, future research based on the finding of this study can (1) investigate an automatic way of risk detection for highway projects using historical data and NLP and (2) measure similarity of risk registers under various features such as project size and location. Further, expansion of the data set promises a more comprehensive set of common components, highlighted by those with frequent and high consequent risks under each project specification. Study limitation include using a pre-trained Word2Vec model. Future studies could increase volume risk registers for comparison and train deep neural networks using the words from a large construction database corpus to improve the model accuracy.

# CHAPTER 3: PREDICTIVE RISK MODELING

The contents of Chapter 3 are published in the Journal of Automation in Construction, Elsevier.

## 3.1 Abstract

Most of the construction practices in the field of risk identification focus on the expertise, views, and judgments of subject matter experts. While the conventional expert-based approaches provide worth, several challenges exist due to time-consuming and expensive aspects. Moreover, limited experience in major projects makes public agencies susceptible to subjective judgment biases. To address these limitations, this study introduced a data-driven framework for risk identification using historical data and artificial intelligence techniques, particularly word embedding models. The model matches various risk items in past projects by considering the semantic meaning of words to find high frequency and consequence risks. Risk registers from more than 70 U.S. major transportation projects form the input dataset. The model is tested with more than 66% recall and 0.59 $F_1$-score for risk detection for new projects. Acquired knowledge from previous projects assists project teams and public agencies to be well-equipped with a risk identification model instead of starting from scratch.

**3.2 Research Design and Data**

3.2.1 Data Collection and Preprocessing

The general expectation is that risks are project-dependent and the unique nature of individual projects drives risk items. However, the authors' research (Erfani et al. 2021b), discussed in chapter 2, demonstrated a huge level of similarity among risk registers in major transportation projects. The risk similarity constructs the foundation of using historical data in this study to propose an initial risk template for major transportation projects.

The main goal of this study is to introduce an NLP-based model that can automate the process of risk register template generation using historical data. This data-driven model considers both frequency and cost/schedule consequences as criteria for developing risk templates. Also, this approach allows risk register customization according to specific project characteristics.

The Information Source for Major Transportation Projects (ISMP) developed by the University of Maryland researchers served as the data source for this study. The database is accessible at https://www.transportationproject.org (Zhang et al. 2022). Containing almost all major highway projects in the U.S. over the last two decades, the database comprises risk registers from 70 projects with more than 6,000 individual risk items. The dataset covers a range of project types, project sizes, delivery methods, and locations to enable the development of reliable risk templates. Figure 8 exhibits the data profile of the collected database. The authors collected risk registers from multiple Excel and pdf files and tabulated results into a comprehensive risk database. While most of the documents follow the standard structure defined by the FHWA including risk name, risk description, risk evaluation, risk response, and allocation, some documents report variations from the standard structure. We defined a comprehensive structure to capture all documents under the same database.

Collected risk registers include the initial risk registers (i.e., ex-ante version) and updated risk registers through the project life-cycle. While the last-updated risk register (i.e., ex-post version) reflects higher data quality by documenting risk items' final status, the authors lacked access to this information for incorporation into this study. The current dataset offers the best-gathered knowledge regarding the risk management practices in major highway projects in the U.S. It provides a valuable summary of diligent efforts by various agencies in the past 20 years to conduct risk studies.

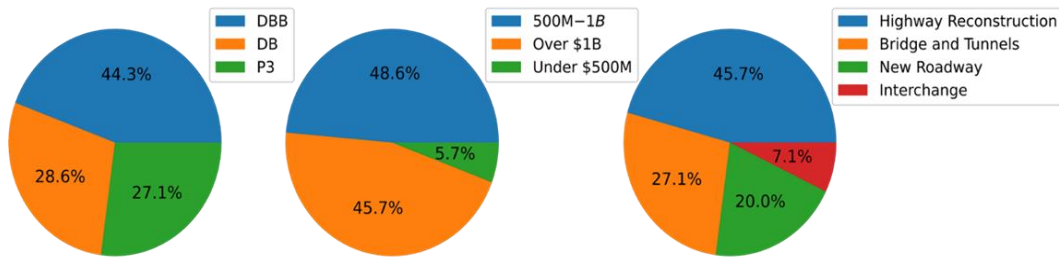

Figure 8. Data profile Summary

3.2.2 Model Development

Flow-Chart

This section presents the methodology of the proposed major transportation projects risk identification model. In the first step, the user selects characteristics that filter projects retrieved from the database. The user could define characteristics such as type, delivery method, size, and location. For all characteristics, an option exists to select all instead of a specific choice. It should be considered that adding all filters could result in decreasing the number of retrieved projects significantly which could result in bias in risk template performance. Then, the model utilizes word embedding models to generate the risk template based on the selected risk registers in the first step. The risk template might be sorted based on prevalence or consequences in terms of cost and schedule. Figure 9 illustrates the main steps of the proposed model, Figure 10 displays the pseudocode, and the following subsections describe the process in further detail.

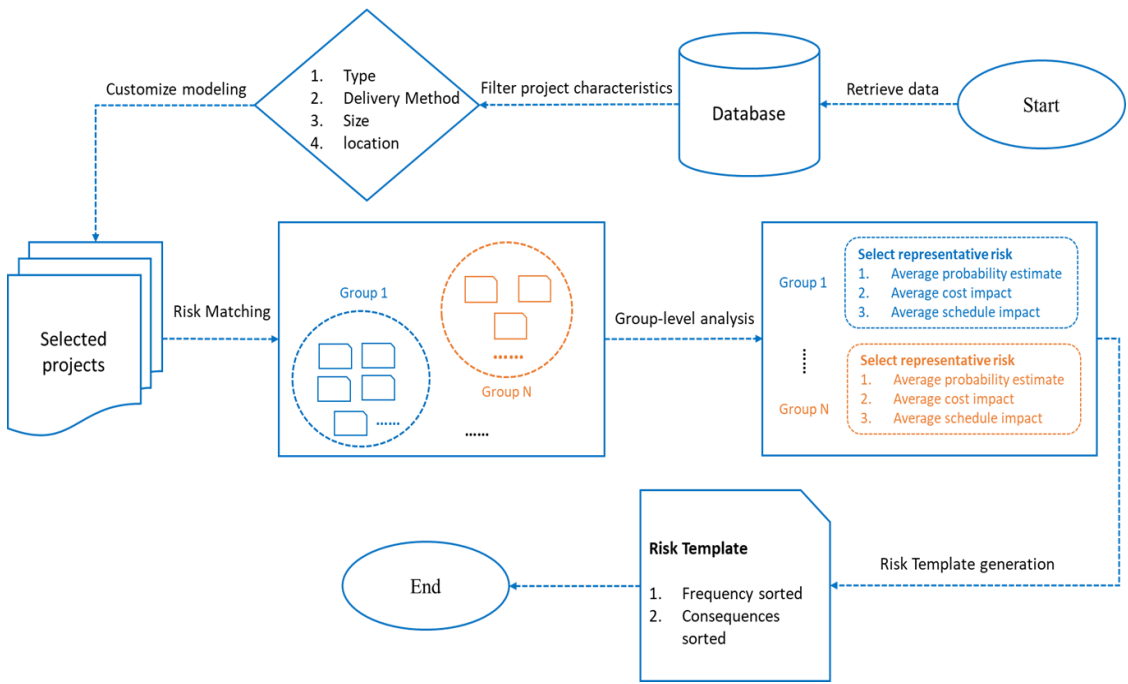

Figure 9. Proposed risk identification model

Step 1: Customize modeling

This model generates a risk register template filtering on user project characteristics preferences. In the first step, we define project characteristics including project type, size, location, and delivery method from the associated provided list. The project type list includes highway reconstruction, bridge and tunnel, new roadway, and interchange. Project sizes subdivide into less than $500M, $500M - $1B, and more than $1B. Project delivery methods consist of design-bid-build, design-build, and public-private partnerships. By defining project characteristics, similar projects with required features are retrieved from the database to generate the initial risk template. The provided flexibility offers users multiple options to generate risk registers. In this way, users might evaluate multiple options of risk templates to customize their risk register accordingly.

| Risk Register Template Generation |
|---|
| **Input:** project type, project size, delivery Method, number of risk items, sorting criteria |
| **Output:** Risk template |
| **Data:** ISMP (70 U.S. major transportation projects) |
| 1: **Function** Customize Modeling: |
| 2: project type= select (all, highway reconstruction, bridge and tunnel, interchange, new roadway) |
| 3: project size= select (all, under $500M, $500M to $1B, over $1B) |
| 4: delivery method= select (all, DB, DBB, P3) |
| 5: number of risk items= select (10, 20, 30) |
| 6: sorting criteria= select (prevalence, cost impact, schedule impact) |
| 7: selected projects= select (project type, project size, delivery method) |
| 8: **Function** Risk matching: |
| 9: Def risk matching (selected projects): |
| 10:     For project in selected projects: |
| 11:         For risk item in project risk register: |
| 12:             Calculate similarity (risk item 1, risk item 2) |
| 13:             If similarity >= 0.7: |
| 14:                 Put risk item 1 and risk item 2 in same group |
| 15: **Function** Group level analysis: |
| 16: Def group level analysis (risk groups): |
| 17:     For risks in the same group: |
| 18:         Risk representative= Frequent language in the group |
| 19:         Risk type= Classify risk type |
| 20:         Avg probability= Avg (risk probability) |
| 21:         Avg cost impact= Avg (risk cost impact) |
| 22:         Avg schedule impact= Avg (risk schedule impact) |
| 23: **Function** Risk register template generation: |
| 24: Def risk register template generation (risk group results): |
| 25:     Risk template = sort group risk using the sorting criteria |

Figure 10. Proposed risk identification model pseudocode

Step 2: Risk matching

Since individual project teams utilize their own language and words to describe the risks, the first step implemented NLP modeling to identify similar risks with varying terminologies. NLP studies convert text into a numeric format for further calculations, called text vectorization (Li et al. 2021; Shahmirzadi et al. 2019).

Data cleaning involves lowercasing, removing stop words, tokenization, and vectorization, creating a list of words. Each word represents a vector, and a risk item is expressed as the average vector of words in a vector space. In order to measure the similarity of two vectors, cosine similarity offers a general measure to calculate the cosine of an angle between two vectors. Eq. 4 displays the mathematic formula to compute the cosine similarity (Sidorov 2019):

$$Similarity(Risk1, Risk2) = Cosine\ (v, w) = \frac{V.W}{||V||||W||} = \frac{\sum_{i=1}^{k} V_i.W_i}{\sqrt{\sum_{i=1}^{k} V_i^2} \times \sqrt{\sum_{i=1}^{k} W_i^2}} \qquad (4)$$

Table 5 provides multiple examples of using the Word2Vec model to calculate the similarity between different risk items. The cosine similarity calculation provides a path to match similar risks with different terminologies from various risk registers to obtain common risks in major transportation projects. A perfect match occurs when the similarity index equals 1, meaning that the same risk item repeats exactly in different projects. To define the threshold of meaningful matching, we selected a random sample of 250 risk matches. Two members of the research team manually read and evaluated the level of similarity among matched risks. Accuracy calculates as the number of accurate matches divided by the total number of samples. While Table 6 indicates that the 60% cosine similarity established an acceptable threshold of similarity to match risks, to keep consistency with the typical practice of considering 70[th] percentile risk-based estimation at FHWA and put a stricter threshold, we set 70% as the threshold of similarity to match risks.

Table 5. Risk matching examples using Word2vec

| Risk (1) | Similarity | Risk (2) |
|---|---|---|
| Owner directed changes and design views | 1 | Owner directed changes and design views |
| Coordination with other projects and with adjacent property owners | 1 | Coordination with other projects and with adjacent property owners |
| Utility Relocations | 1 | Utility Relocations |
| Determination of secondary impacts to wetlands | 0.95 | Determination of wetlands impacts |
| Asbestos and Lead Paint | 0.94 | Environmental issues - asbestos or lead paint |
| Utility Relocations | 0.85 | Utility Relocations and conflicts |
| Federal agencies may take longer than expected to review and issue a permit | 0.81 | Permits or agency actions delayed or take longer than expected |
| Handling of Contaminated Materials | 0.78 | Unanticipated Hazardous Materials or Contaminated Soils |
| Unforeseen Utilities | 0.72 | Unknown Utilities |
| Construction and utility costs | 0.64 | Lack of general maintenance during construction |
| Extend time frame Proposer Request for Proposal Responses | 0.62 | Utility agreements prior to final request for proposal |
| Delay Claims | 0.59 | Delay in agreement |
| Hazardous materials | 0.54 | Quality and availability of equipment, materials and labor |

Table 6. Similarity matching threshold selection

| Cosine similarity level (%) | Sample size | Accuracy (%) |
|---|---|---|
| 100 | 50 | 100 |
| 80-100 | 50 | 100 |
| 70-80 | 50 | 86 |
| 60-70 | 50 | 80 |
| 50- 60 | 50 | 60 |

A Python script exploits the word embedding model and performs automatic cosine similarity calculation. More than 6,000 risk items from 70 projects reside in the database. Code compares the first risk from the first selected project against all items from retrieved projects based on user preferences. If Risk 1 reaches 70% similarity to any risk in the database, they are classified into the same group. In the second loop, matches from previous loops are excluded, and the process continues until all items in the selected projects find their match. The result of this step returns a list of grouped risks with different wording and unique assessment by various project teams

Group-level analysis

The next analysis evaluates previously grouped risks, to find the text that represents the group and calculate the average probability of occurrence and consequences in terms of cost and schedule impact. Sometimes, different project teams use the same verbiage to describe risks. In each group, the text that was repeated more frequently by the project teams serves as the group representative. Results allow calculation of the average probability and cost and schedule impact. Table 7 provides an example of similar risk items with different terminologies in DBB projects and how the group-level analysis automatically finds the final risk item. As 27 out of 31 DBB projects include a risk related to the timely utility relocation the prevalence of this risk equals 87%. "construction impacts due to lack of right of way and timely utility relocation" will be selected as the representative risk, due to more usage by various project teams. Note risk items residing in the same group report at least 70% semantic cosine similarity based on their text to be considered under the same group.

Table 7. Group-level analysis example

| Risk item | Frequency (out of 27) |
|---|---|
| construction impacts due to lack of right of way and timely utility relocation | 7 |
| right of way acquisition needed prior to utility relocations and construction | 5 |
| utility relocation at overcrossings | 4 |
| utility relocation may not happen on time | 3 |
| utility relocation may not happen in time | 3 |
| delays in utility relocations due to delay in field work required from utility owners | 2 |
| global risk related to utilities relocation | 2 |
| encounter unanticipated and unknown utilities or damage to utilities during construction | 1 |

Furthermore, to generate a structured risk identification result, researchers utilized a risk classifier to detect the best risk type for each item. Authors classified the identified risks into ten groups including "environmental", "structure and geotechnical", "design", "right of way", "utilities", "railroad", "partnerships and stakeholders", "management and funding", "contracting and procurement", and "construction" based on Washington Department of Transportation (WSDOT) risk breakdown structure. The authors first removed the stop words (e.g., the, is, and, but). Second, programming tokenizes each risk item into a list of words. After applying the same process for risk categories, the cosine similarity between risk names and risk categories was calculated. Each risk item was assigned to the group with the highest text similarity. Table 8 provides some examples of how the risk type for risk items was detected. While one risk item might relate to the multiple risk type groups and not easily be considered under one group, the purpose of the classification detects the highest similar group. For example, the risk item "potential changes to geotechnical design for foundation" contributes to both aspects of "structure and geotechnical" and "design". Indeed, the similarity calculation ranked "structure and geotechnical" and "design" first and second respectively. The risk is assigned to the top similar group.

Table 8. Risk classification examples

| Risk item | Risk categories | Similarity | Selected type |
|---|---|---|---|
| Additional right of way required | Environmental | 0.132 | Right of way |
| | Structure and geotechnical | 0.231 | |
| | Design | 0.162 | |
| | Right of way | 0.778 | |
| | Utilities | 0.222 | |
| | Railroad | 0.108 | |
| | Partnerships and stakeholders | 0.176 | |
| | Management and funding | 0.263 | |
| | Contracting and procurement | 0.171 | |
| | Construction | 0.161 | |
| Potential changes to geotechnical design for foundations | Environmental | 0.442 | Structure and geotechnical |
| | Structure and geotechnical | 0.797 | |
| | Design | 0.582 | |
| | Right of way | 0.201 | |
| | Utilities | 0.281 | |
| | Railroad | 0.200 | |
| | Partnerships and stakeholders | 0.326 | |
| | Management and funding | 0.376 | |
| | Contracting and procurement | 0.240 | |
| | Construction | 0.479 | |

Step 4: Risk register template generation

By grouping risks and computing the average consequences for each group, a variety of potential methods customize templates. Frequency in actual projects, average probability, or average cost and schedule consequences drive sorting and prioritization. Therefore, risk register development requires answering two main questions. First, what criteria should be applied for sorting, and second, how many unique risk items should be considered in the template? Multiple options can be considered in the model to generate a risk template based on user preference. Users generate templates based on risk frequency or consequences. Also, justified by FHWA guidelines, an option exists to include the top 10-20-30 items in the risk template. No numerical limits exist in the template and the choice depends on the specific application. Template length should adequately consider all major risks while avoiding excessive detail to distract the project team focus from critical risk items. Therefore, 10 to 30 risk items in the risk register seem adequate.

**3.3 Results and Discussion**

3.3.1 Model testing and validation

Word embedding model selection

The main part of the proposed predictive risk model depends on the deployed word embedding technique because it plays an important role to find similar risks with different languages. Authors selected the best NLP approach to convert text to vectors to perform the study. A detailed experiment was conducted among NLP models (e.g., Word2vec, FastText). To set up the testing process, we collected the risk registers from five new projects excluded from the main database. Table 9 provides detailed information about the testing projects.

Table 9. Testing projects detailed information

| Project ID | Jurisdiction | Delivery Method | Project Type | Project size ($) | Risk items |
| --- | --- | --- | --- | --- | --- |
| A | FL | DB | Bridge and Tunnel | 500 M - 1 B | 35 |
| B | CA | DBB | Highway reconstruction | More than 1 B | 38 |
| C | TX | P3 | Highway reconstruction | More than 1 B | 50 |
| D | NJ | DB | New roadway | 500 M - 1 B | 13 |
| E | WV | DBB | Interchange | 500 M - 1 B | 7 |

In the next step, each risk item for each testing project was compared to all of more than 6,000 risk items in the database and matched to the highest similar risk based on the cosine similarity calculation. The process was repeated using word embedding models including Word2Vec, FastText, and GloVe. The authors observed that while these three models found the same risks in many cases, in some cases the highest similar risk returned differently for different models. The question is which model is more intelligent to find better risks to match.

Further action includes the human labeling process. Two members of the research team manually read the matched risks, dividing them into three groups of "high", "medium", and "low" similarity. High similarity represents risk items containing the same topic and application. Medium similarity infers that the risks share a similar concept but have different applications, while low similarity represents completely different risks. Each member cross-checked the matched risks and upon receiving the same label from both members, items were considered final. Otherwise, both team members read the risks together and selected the final label through discussion. Table 10 offers some examples of similarity under each group of "high", "medium", and "low".

Table 10. Risk labeling examples

| Risk (1) | Similarity level | Risk (2) |
|---|---|---|
| Coordination with other projects and with adjacent property owners | High | Coordination with other projects and with adjacent property owners |
| Hazardous materials | High | Handling of contaminated materials |
| Federal agencies may take longer than expected to review and issue a permit | High | Permits or agency actions delayed or take longer than expected |
| Different site conditions | Medium | Encountering additional Archaeological Sites |
| Oil lines Relocation and Right of Way | Medium | Reduce right of way corridor (outside the lanes) width in some areas |
| Delay in agreements | Medium | Breach of obligations and agreements by private sector |
| Structure Opportunity | Low | Additional right of way required |
| Utility risk | Low | Structural steel price escalation |

Two main criteria need to be considered for the final model selection. First, the model should be intelligent, in other words, find the highest similar risk appropriately. Second, the model's index should be accurately reflecting the level of similarity. In other words, the decreasing similarity between matched risks results in proportionally lower cosine similarity. Table 11 reveals that the Word2Vec model significantly outperforms the other two options over the same data points in terms of finding the highest similar risk.

Table 11. Word embedding models' performance comparison

| Level of similarity | Word2Vec | FastText | Glove |
|---|---|---|---|
| High | 81.1% | 66.9% | 63.5% |
| Medium | 9.5% | 12.2% | 10.8% |
| Low | 9.5% | 20.9% | 25.7% |

According to the authors' observation, not only did the Word2Vec match risks more accurately but also the similarity index better reflected the similarity level. Among multiple risk item examples in Table 10, the similarity index range for Word2Vec appeared more accurate than the other two models. Therefore, researchers selected Word2Vec to be deployed in this study. Furthermore, for more than 81.1% of risk items in five testing projects, at least, one project in the database reported the same risk in their risk registers. It should be noted that 60% of high similarity matched risks even use the same language and verbiage. The results show a promising application of historical risk data to support risk analysis on new projects. This data-driven approach allows project teams to concentrate on project-specific risks by quickly offering common risk templates based on a comparison of early projects.

**Risk register template performance**

The purpose of the next experiment was to evaluate the performance of the developed risk template by using the initial five projects. This process included the following steps:

1. Generate a risk template containing an intended number of risks, for instance, 30 items
2. Find the closest match for each one of the risks in testing projects to the risk template
3. Manually cross-check to assess whether the match was high, medium, or low
4. Score the risk template performance using metrics such as recall, precision, and $F_1$-score

While we did not expect that all the risk items were covered in a short risk template, we anticipated that the common risks were detected by the developed risk template. To calculate the performance metrics, various outcomes need to be clarified. True-positive (TP) reflects risk items that exist in both the testing risk register and developed risk template (i.e., matched risks with high or medium similarity levels). False-negative (FN) occurs when risk registers items fail to appear in the template (i.e., matched risks with a low similarity label). Also, the False-positive (FP) contains the risks included in the risk template but has not been used for matching any risks in the testing risk register. Equation 5 calculates the performance metrics of the risk template.

$$\text{Recall} = \frac{TP}{TP + FN}; \text{Precision} = \frac{TP}{TP + FP}; F_1 = \frac{TP}{TP + \frac{1}{2}(FN + FP)} \quad (5)$$

Table 12 and 13 depicts the results for testing projects compared to the general template with all 70 projects. Results include no definition of specific project features, including 30 risk items in the template, and are sorted based on the prevalence in previous projects.

Table 12. Risk matching outcomes

| Project ID | High | Medium | Low | TP | FN | FP |
|---|---|---|---|---|---|---|
| A | 4 | 12 | 19 | 16 | 19 | 19 |
| B | 12 | 9 | 17 | 21 | 17 | 19 |
| C | 32 | 9 | 9 | 41 | 9 | 2 |
| D | 9 | 3 | 1 | 12 | 1 | 18 |
| E | 1 | 4 | 2 | 5 | 2 | 25 |
| Overall | 58 | 37 | 48 | 95 | 48 | 83 |

Among the metrics, recall reflects the performance of the risk template better due to a high cost associated with false negatives (i.e., risks from the testing project that fail to be included in the template). Precision illustrates what portion of risks in the template is identified correctly. F1-score seeks a balance between precision and recall. According to the results, a simple and short risk template using objective historical data averages reports more than 66% recall. Results range from 92.3% to 45.7% of risk items covered by the risk template. It is expected that if the project team utilizes specific language or considers more unique risks the risk template performance decreases. The results suggest a great potential for a common data-driven risk template to initiate the preparation of risk registers for major transportation projects overall. In the following experiments, the authors applied tighter criteria, evaluated the impact of sorting the risk template based on the consequences, and added more specific project features on risk template performance.

Table 13. Risk template performance

| Project ID | Recall | Precision | $F_1$-score |
|---|---|---|---|
| A | 45.7% | 45.7% | 45.7% |
| B | 55.3% | 52.5% | 53.8% |
| C | 82.0% | 95.3% | 88.2% |
| D | 92.3 % | 40.0% | 55.8% |
| E | 71.4 % | 16.7% | 27.0% |
| Overall | 66.4% | 53.4% | 59.2% |

**Prevalence vs consequences**

The proposed data-driven model captures not only the critical risks based on the frequency of occurrence in actual historical project data but also the potential consequences in terms of cost and schedule. Therefore, in this sub-section, researchers repeated the previous experiment, but sorted the risk template based on cost and schedule impact. To do so, the risk group items are sorted based on average consequences calculated in the model based on previous similar projects. According to the result (Table 14), risk template performance for the prevalence risk template outperformed the consequence-based (i.e., cost and schedule impact risk templates) by more than 10% in terms of recall and F1-score. However, the flexibility which is provided to sort the risk template based on the high consequent risks might be considered a valuable option provided by analyzing historical data.

Table 14. Prevalence and consequence-based risk template comparison

| Project ID | Recall | | | $F_1$-score | | |
|---|---|---|---|---|---|---|
| | Prevalence | Cost impact | Schedule impact | Prevalence | Cost impact | Schedule impact |
| A | 45.7% | 18.1% | 20.0% | 45.7% | 18.5% | 21.2% |
| B | 55.3% | 34.2% | 48.6% | 53.8% | 38.2% | 45.9% |
| C | 82.0% | 68.0% | 70.0% | 88.2% | 69.4% | 73.7% |
| D | 92.3% | 84.6% | 69.2% | 55.8% | 47.8% | 41.9% |
| E | 71.4% | 71.4% | 42.9% | 27.0% | 27.0% | 16.2% |
| Overall | 66.4% | 55.3% | 50.1% | 59.2% | 43.9% | 45.1% |

**Sensitivity analysis**

Key project feature features include type, delivery method, size, and location to identify previous projects with matching attributes. While there are other potential drivers such as site condition behind project risk selection, in this sub-section, researchers evaluated the risk template performance developed based on adding each characteristic of the testing projects. For each testing project, we retrieved cases that match the selected characteristic and developed the risk template following the proposed model. For example, if project A is located in the state of Florida, all projects in the database from that agency are retrieved to develop the risk template. Then a similar experiment to the previous sub-sections evaluated the risk template performance. Table 15 illustrates how the overall risk template performance has been changed in comparison to the developed risk template in subsection 6.2 as the basis. According to the result, the project location significantly helps to improve the risk template performance in terms of recall, precision, and F1 score (an increase of 9.1%, 3.7%, and 5.9% respectively). The potential reason behind the observation is that public agencies typically work with the same consultants for major projects and those consultants build their knowledge through their previous experiences with risk studies. This observation emphasizes the importance of data quality in terms of replicating good practices and the danger of modeling previous projects' mistakes. Other features such as project type and size report small improvements (2.1% and 0.7% increase in recall respectively). Also, choosing the delivery method failed to improve the risk template performance. Ultimately, the new project situation and project team evaluation determine the final decision. The proposed model provides suitable flexibility to consider various options to generate an initial risk template based on various project characteristics.

Table 15. Project characteristics selection impact on risk template performance

| Project characteristic | % of change in Recall | % of change in Precision | % change in $F_1$-score |
|---|---|---|---|
| Type | + 2.1% | + 0.2% | + 0.9% |
| Delivery Method | - 1.4% | -1.7% | - 1.6% |
| Size | + 0.7% | + 0.2% | + 0.4% |
| Location | + 9.1% | +3.7% | + 5.9% |

3.3.2 Discussion

The proposed predictive risk model capitalizes on historical data to produce risk templates for major transportation projects. Various advantages of this data-driven risk identification technique exist in comparison to expert judgment-based approaches. First of all, this approach provides significant flexibility for the customization of the risk template based on various project characteristics. It also enables the user to consider cost/schedule consequences besides the frequency of risk items in historical data. Moreover, the proposed model requires less time and cost compared to traditional risk identification approaches. One of the main challenges of the subject matter experience-based approach is subjectivity. Therefore, expert bias significantly influences the risk identification results. However, the data-driven approach expedites processing while capturing the subjectivity uncertainty of experts' judgments using empirical data.

Finally, the proposed model provides easy transmission of lessons learned from similar previous projects in terms of the risks and average consequences. The flexibility of this approach allows reuse for other project types with tabulated databases. In addition to all the theoretical contributions of this paper, the main benefit of the proposed model focused on its application in real highway projects. Project teams benefit from a tool that expedites risk identification, reducing the time required to identify common risks. Noted the testing experience and sensitivity analysis underline the variation in model performance in different scenarios. Specifically, to implement the model, the potential bias in those scenarios in which the number of data points is low should be considered. But, the proposed tool is transparent in terms of sample size and accuracy. It plays as an advisory tool to help project teams as an initial step in conducting risk studies.

One of the main limitations of the current study is the lack of access to ex-post risk data. Ex-post data would reflect what actually happened by project completion. That would significantly increase the value of the current study by considering the real values of delays and cost overruns besides how successful the project was. It should be noted that the developed model in this study could be calibrated and applied to a new dataset of the ex-post risk data.

## 3.4 Conclusion

This study offered predictive risk modeling using historical data from major transportation projects instead of relying on expert judgment solely. This data-driven approach utilizes NLP and word embedding models to detect similar risks with various terminologies from risk registers of similar previous projects. The model is able to capture critical risks considering both prevalence and cost/time consequences as well as specific project characteristics.

The authors tested the application of the proposed model by evaluating the risk template over five sample projects. Results revealed that the model aids project teams in automatically detecting more than half of the common risk register items. Testing experiments suggest that adding project characteristics such as location, type (e.g., bridge, highway, interchange) and size improve the risk template performance.

Results suggested that upon the high similarity in risk items for major transportation projects, the proposed model can offer an initial step in conducting risk studies to help project teams become equipped with knowledge and experience of similar previous projects. Project team judgment refines and finalizes the risk template in accordance with the unique nature of each project. Limitations to the proposed solution require future studies, highlighted by examination of the ex-post risk data incorporating the actual impact based on what happened and response application for new projects. Also, study limitations include a relatively small sample size and using a pre-trained Word2Vec model. Future studies could increase volume risk registers for comparison and train deep neural networks using the words from a large construction database corpus.

# CHAPTER 4: *EX POST* PROJECT RISK ASSESSMENT

The contents of Chapter 4 are published in the Journal of Construction Engineering and Management, ASCE.

**Citation**: Erfani, A., Ma, Z., Cui, Q., Baecher, G. (2023b). Ex post Project Risk Assessment: Method, and Empirical Study, *Journal of Construction Engineering and Management*, 149(2), 04022174.

## 4.1 Abstract

Project risk is an important part of managing large projects of any sort. The study contributes to the state of knowledge in project risk management by introducing a data-driven approach to measure risk identification performance using historical data. In the early phases of a project, the identification and assessment of risk is based largely on experience and expert judgment. As a project moves through its life cycle, these identified risks and their assessments evolve. Some risks are realized to become issues, some are mitigated, and some are retired as no longer important. The study investigates the quality of early risk registers and risk assessments on large transportation projects in comparison to how those risks evolved on historical projects. It does so by using textual analysis of archival risk registers documents. Finite state automation methods akin to Markov Chain models are used to track the changes in risk attributes on these large infra-structure projects as the projects mature. The objective is to be better able to anticipate how such project risks will change as projects move forward and to be better able to forecast changes to the risk register from ex ante to ex post conditions.

Results from 11 major US transportation projects suggest that on average somewhat fewer than 65% of ex ante identified risks ultimately occur in projects and are mitigated, while somewhat more than 35% do not occur and are retired. In addition, more than half of the risks emerged during project execution when new information became available. Categorizing risk management styles illustrates that planning for identified risks in the initial phase of the project is necessary but not sufficient for successful project delivery. A project team with positive doer behavior (i.e., actively monitoring and identifying risks during project execution) performed better in delivering projects on time and within budget.

**4.2 Research Design and Data**

Finite state automation methods, akin to Markov Chain models, are used to track changes in risk attributes as a project matures. The objective is to be better able to anticipate how project risks change as a project moves forward and to be better able to forecast changes to the risk register from *ex ante* to *ex post* conditions.

A finite-state automaton (FSA) is a simple computational model which defines a list of states a system can occupy and permissible transitions among those states. The system can be in one and only one state at a given time (the status of the system). The FSA transitions among states in response to some inputs. An FSA is fully specified by its set of states, an initial state, and the inputs that prompt transitions. The final state of the system is an output of the initial state and a series of inputs. The common representation of an FSA is a state-transition table. This indicates, for each state and input, what the next state will be. An FSA can be either deterministic or probabilistic.

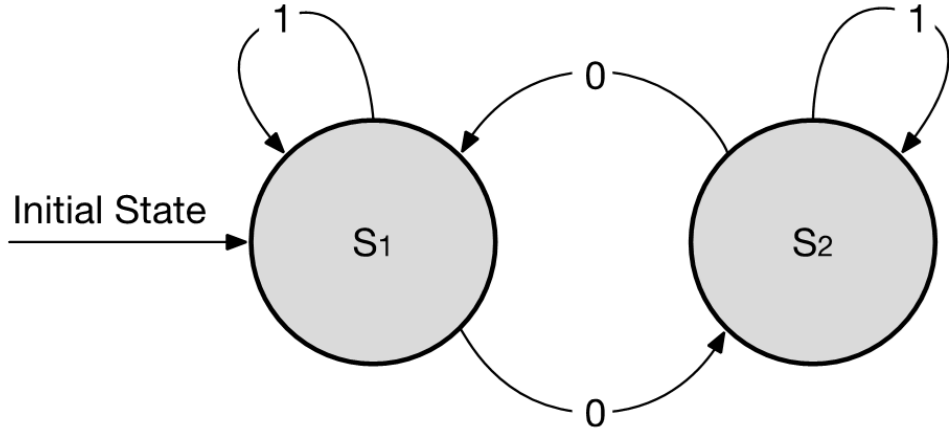

Figure 11. Simple two-state system with two inputs

The simple system of Figure 11 has two states, $Q = (S_1, S_2)$, and two possible inputs, $\Sigma = (0,1)$. The state-transition table is given by Table 16. The system state changes if the input is 0 and remains the same if the input is 1. These transitions are represented as, $\delta = (continue, change)$. Thus, using standard FSA notation, the system can be represented as,

$$\text{FSA} = (Q, \Sigma, \delta, q_0, F) \qquad (6)$$

where,

$Q$ = a finite set of states

$\Sigma$ = A finite set of inputs

$\delta$ = a transition function, $Q \times \Sigma \to Q$

$q_0$ = an initial state, $q_0 \in Q$

$F$ = set of final states, $F \in Q$

The FSA starts at one initial state and ends at a final set of states resulting from the sequence of inputs.

Table 16. State-transition table

| Input | Current state | |
|---|---|---|
| | $S_1$ | $S_2$ |
| 1 | $S_1$ | $S_2$ |
| 0 | $S_2$ | $S_1$ |

The FSA computational model is widely used in computer science (Bahattacharya and Ray 2022; Shannon 1953; Vardi 1989), mathematics (Mackenzie 1995; Wen and Ray 2012), biology (Chen and Mynett 2003; Ermentrout and Edelstein-Keshet 1993), linguistics (de Case et al. 2009, 2012), and engineering (Freire and DaCamara 2019; James 2019; Richter et al. 1999; Song et al. 2021). In recent years, applications have appeared in the project management literature. Shimura and Nishinari (2014) applied Cellular Automata, an extension of FSA, to network scheduling. Flood and Goodenough (2021) used cellular automata to represent the contracts in a computational format. Anari et al. (2013) used learning automata, another extension of FSA to optimize the risk management process. Baiardi et al. (2008) implement FSA to understanding sequences of complex cyber security risks.

### 4.2.1 Risk life-cycle framework

In construction, various risks arise at different stages of a project's life cycle and evolve dynamically. Risk identification refers to systematically and continuously identifying potential risks and their consequences on a project (Erfani et al. 2022; Siraj and Fayek 2019). Risk identification is conducted by project teams using a variety of tools and techniques during planning and construction. Documentation reviews, information gathering techniques, checklists, and expert judgment, are some of these tools (Al-Al-Bahar and Crandall 1990; Iqbal et al. 2015). Moreover, federal highway and department of transportation agencies have developed various risk identification guidelines and standards for transportation construction projects (Curtis and Program 2012, Molenaar 2006, 2010). A risk register documents the outcome of the risk identification and updates it as the project progresses. It is difficult to evaluate the interaction of risk factors in current risk management practices. Therefore, risk factors are identified independently and analyzed separately (Tavakolan and Etemadinia 2017).

Project risks can be tracked through different states, transforming from one to another as the project progresses. Using the automata theory concept (O'Regan 2021), risk state refers to the risk status at various stages in a project's lifecycle and risk transition functions govern the transition between states.

**Risk states,** In the present study, risk states are defined as *registered*, *happening*, and *closed* (Table 17). A risk is classified as *registered* once it has been identified. It transitions to *happening* when it begins impacting the project. During project execution, some risks might be realized (*i.e.,* occur) and some risks might be dismissed (*i.e.,* remain dormant or have no impact). These risks states are designated, *closed*.

Table 17. Risk states and definitions

| Code | Risk State | Definition |
|------|------------|------------|
| Reg | Registered | The risk item is identified, generated, and not start to happen |
| Hap | Happening | The risk item is currently occurring and hasn't been closed yet |
| Clo | Closed | The risk item close after/without occurrence |

**Risk transitions,** Risk states identified *ex ante* may change during project execution and new risks may emerge. The transition function defines how risks evolve (*i.e.,* move from one state to another). The transitions are categorized in four types: *generate, occur, continue,* and *close* (Table 18). In the symbology of FSA, these are represented as, $\delta = (generate, occur, continue, close)$.

Table 18. Risk transition functions and definitions. $\emptyset$ indicates an initial state of before the risk is formally identified.

| Risk Transition | Definition | State transition |
| --- | --- | --- |
| Generate | The risk item is generated and added to the risk register | $\emptyset \rightarrow$ Reg |
| Occur | The risk item happens | Reg $\rightarrow$ Hap |
| Continue | The risk item continues the state | Reg $\rightarrow$ Reg <br> Hap $\rightarrow$ Hap <br> Clo $\rightarrow$ Clo |
| Close | The risk item is closed | Reg $\rightarrow$ Clo <br> Hap $\rightarrow$ Clo |

*Generate* refers to when a new risk identified during construction, *occur* refers to a risk that starts to happen. *Continue* refers to a risk that remains in the same state as the project moves forward. *Close* indicates refers to a risk has been mitigated or dismissed. All risks are classified as closed at project completion.

**Risk lifecycle.** Each risk can have a unique lifecycle from the time that it has been initiated to the time it is closed. A *Risk Life-cycle Automaton* (RLA) can be represented as in Figure 12. The RLA includes the three risk states of registered, happening, or closed; and the four transitions of generate, continue, occur, and close.

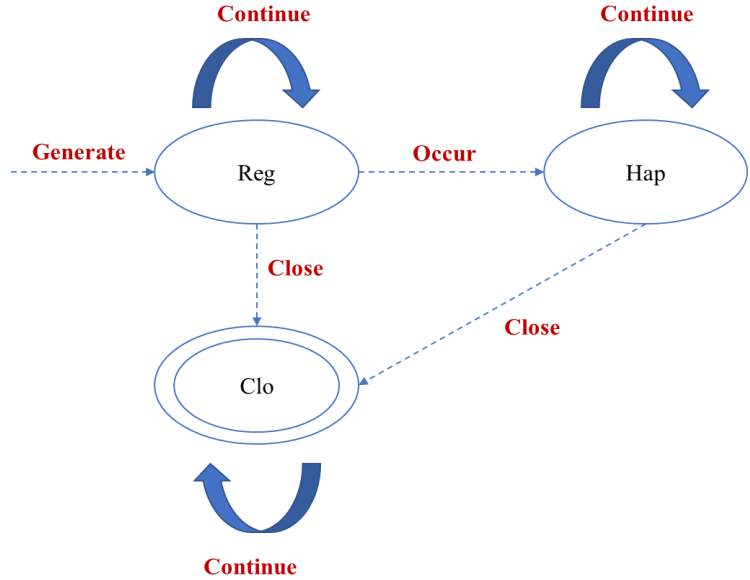

Figure 12. Risk life-cycle automaton (RLA)

The final state, marked with a double circle, will be reached if when the risk is closed, which occurs for all risks at the end of the project. In this model, the risk states are executed based on the set of operations the risk transitions enable, that is, the set of transitions for which their preconditions hold. Hence, using automaton characterization, the RLA can be written as,

$$\text{FSA} = (Q, \Sigma, \delta, q_0, F) \tag{7}$$

where,

$Q = $ (Reg, Hap, Clo)

$\Sigma = $ information accruing during construction

$\delta = $ (occur, continue, close)

$q_0 = $ (Reg)

$F = $ (Clo)

Given the RLA, each risk is stored by the sequence of state combinations it follows. The RLA automated provides a clear and intuitive interpretation of a risk life-cycle. Figure 13 illustrates three examples of possible risk life-cycles.

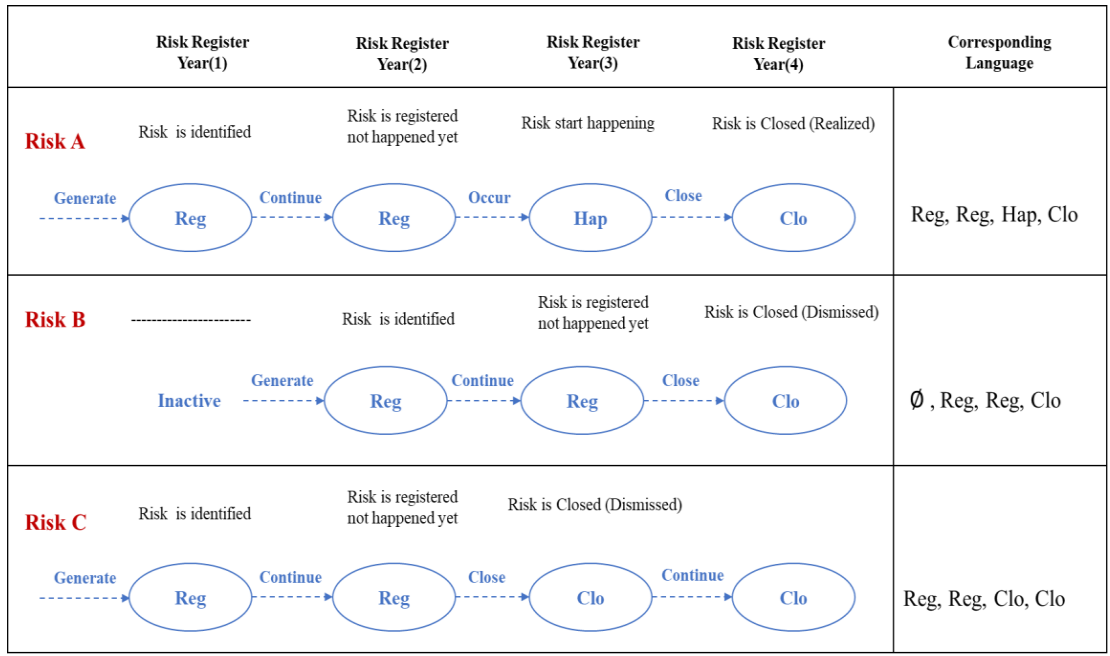

Figure 13. Risk life-cycle automata examples

4.2.2 Procedure

Ultimately, the goal was to measure risk management performance by risk identification and following risks from either their *ex ante* or during-project-execution beginnings to their *ex post* conclusions (which will always be Clo eventually). The approach was to use the proposed framework to follow evolving (or emerging) risk states through the project life cycle. This can be done by tracking the proportion of identified risks that is realized by the project's end and the portion that is dismissed. Utilizing historical risk registers, the first step includes applying the proposed framework to identify individual risk life cycle, organize risks into two major categories of initial and construction stage risks, and determine whether the risk occurred or was dismissed.

The proposed framework was applied to 11 major US transportation projects. Table 19 summarizes the project data. Risk registers developed by the project team were extracted from the annual financial updates document as inputs for our analysis. For each project, risk transitions were recorded among the project risk registers by tracking risk items (Figure 14). The process of RLA tracking was based on available data for each project, including risk descriptions, risk response strategies, and their respective likelihoods. For instance, a high occurrence probability (90%-100%) risk that accompanies a cost overturn will be classified to a Happening state in that year. All risks are considered as closed when the project is completed. Even though some risks were recorded as "Registered" status in the last updated risk register document, those risks are considered as "Closed" in the final year.

Table 19. Ex post project dataset and characteristics.

| Project ID | Project Type | Jurisdiction | Delivery Method | Project Size (M $) | Number of risk registers | Number of risks in initial stage | realized Initial risks | Number of risks in construction stage | Realized Construction risks |
|---|---|---|---|---|---|---|---|---|---|
| 1 | Highway | CA | DB | 1421 | 5 | 32 | 31 | 6 | 6 |
| 2 | Highway | IA | DBB | 1131 | 4 | 24 | 21 | 22 | 22 |
| 3 | Highway | TX | DBB | 4922 | 4 | 85 | 72 | 16 | 16 |
| 4 | Highway | CA | DBB | 1792 | 4 | 43 | 39 | 103 | 68 |
| 5 | Highway | CA | DBB | 986 | 4 | 19 | 15 | 28 | 17 |
| 6 | Highway | FL | DBB | 684 | 5 | 131 | 24 | 193 | 188 |
| 7 | Bridge and Tunnel | CA | DB | 1492 | 4 | 65 | 36 | 24 | 9 |
| 8 | Highway | MD | DBB | 814 | 2 | 15 | 9 | 30 | 11 |
| 9 | Bridge and Tunnel | KY | DBB | 583 | 2 | 15 | 3 | 1 | 0 |
| 10 | Highway | TX | DB | 693 | 2 | 15 | 3 | 2 | 0 |
| 11 | Highway | MI | P3 | 1137 | 2 | 14 | 4 | 41 | 3 |

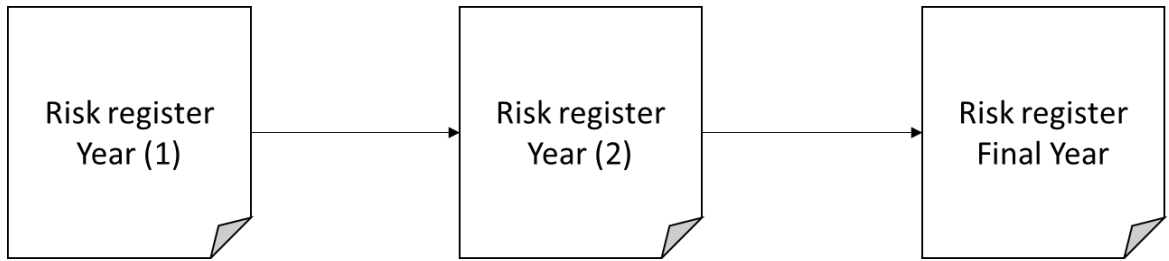

Figure 14. Project data and tabulation process.

Sometimes recorded project data vary in structure and content because each project team may have a unique approach to keeping those data updated. Therefore, dealing with these unstructured data required reasonable assumptions. These were kept consistent for all the projects. These assumptions were:

1) The study assumed that those intermediate "Happening" risks would be considered as realized in the final year if no other information was provided.

2) The study assumed that those intermediate "Registered" risks would be considered as dismissed in the final year if no other information was provided.

3) The study assumed that all risks in the initial version should be "Registered" or "Happening". Risks could not be considered as close in the initial step.

4) The study assumed that all risks in targeted projects were closed in the last updated risk register.

After examining each risk path, a risk was classified as realized or happened if it reached the "Happening" state. In other words, the language corresponding to the Automata language includes a state called "Happening". Risk A in Figure 13, for example, represents a realized risk identified in the initial stage. The risk would otherwise be dismissed and not happen. That means that the corresponding language will never reach the "Happening" state in the automata framework. Figure 13 illustrates two dismissed risks, B and C. Risk B was identified during project execution, whereas risk C was identified at the beginning of the process.

**4.3 Results and Discussion**

4.3.1 Risk identification performance ratios

Proposed indicators (performance ratios) by which to evaluate risk identification performance are shown in Table 20. With the established RLA datasets and classifying initial/construction risks to realized or dismissed categories, the ratios were automatically generated by predefined functions. As an example, in Project ID =4, 43 risks were specified in the initial risk register. 39 of those risks actually happened and the rest were dismissed. According to historical records, the project team added 103 risks during construction, and 68 of these risks eventually occurred (See Table 19). Main performance ratios are calculated as follow:

$$\text{Initial realization ratio} = \frac{\text{Number of realized risks in year 1}}{\text{Number of identified risks in year 1}} = \frac{39}{43} = 0.91 \quad (8)$$

$$\text{Further realized ratio} = \frac{\text{Number of realized risks from risks after year 1}}{\text{Number of identified risks after year 1}} = \frac{68}{103} = 0.66 \quad (9)$$

$$\text{New item ratio} = \frac{\text{Number of identified risks after year 1}}{\text{Total Number of identified risks}} = \frac{103}{43 + 103} = 0.71 \quad (10)$$

$$\text{Total realization ratio} = \frac{\text{Number of realized risks}}{\text{Number of identified risks}} = \frac{39 + 68}{43 + 103} = 0.73 \quad (11)$$

$$\text{Initial efficiency ratio} = \frac{\text{Number of realized risks in year 1}}{\text{Number of identified risks}} = \frac{39}{39 + 68} = 0.36 \quad (12)$$

Table 20. Risk performance ratio definitions

| Scope | Ratio | Ratio Formula |
|---|---|---|
| Overall performance | Total realization ratio | $\dfrac{\text{Number of realized risks}}{\text{Number of identified risks}}$ |
| | Total dismissed ratio | $\dfrac{\text{Number of dismissed risks}}{\text{Number of identified risks}}$ |
| Initial performance | Initial realization ratio | $\dfrac{\text{Number of realized risks in year 1}}{\text{Number of identified risks in year 1}}$ |
| | Initial dismissed ratio | $\dfrac{\text{Number of dismissed risks in year 1}}{\text{Number of identified risks in year 1}}$ |
| | Initial efficincy ratio | $\dfrac{\text{Number of realized risks in year 1}}{\text{Number of realized risks}}$ |
| Construction phase performance | New item ratio | $\dfrac{\text{Number of identified risks after year 1}}{\text{Total Number of identified risks}}$ |
| | Further realized ratio | $\dfrac{\text{Number of realized risks from risks after year 1}}{\text{Number of identified risks after year 1}}$ |

The suggested performance ratios consider three phases of risk identification performance during the project life: overall, initial, and construction phase. First, to evaluate the overall risk identification performance, the "total realization ratio" is measured as a fraction of the number of total realized risks divided by the total of identified risks. The "total dismissed ratio" is calculated from the other side of risk management performance through the dismissed numbers. The sum of these two metrics must equal one.

Risk planning is essential in the initial project phase. In the literature, this is called proactive risk management. The project team tries to identify most of the risks in the early phase of the project (Kaliprasad 2006). To capture this initial planning, the present study computed the proportions of realized and dismissed initial risks through the "initial efficiency ratio," "initial realization ratio" and "initial dismissed ratio." These focus on the project's initial year based on the number of realized or dismissed risks only in the first year.

Regardless of initial risk identification performance, risk monitoring and planning during project execution is essential. This reactive behavior which is sometimes considered as planning for risks that are newly identified or start happening during project execution (Pavlak 2004). To analyze the project risk management performance during execution, the study calculates the percentage of new risks over total risks, named "new item ratio." Another indicator, the "further realized ratio" was considered to estimate the effectiveness of responses for the group of new risks. Note that, the proposed framework could be applied in future studies to track risk assessment performance also with respect to cost and schedule consequences compared to actual values. The cur-rent study only tracks metrics related to risk occurrence frequency.

The average risk identification realization ratio was 0.64. This indicates that 64% of identified risks on average in these testing projects happened during the project life-cycle. That is, 36% of identified risks did not occur. Thus, the total dismissed ratio was 0.36. The performance variation among different project teams was significant. Projects were sorted on the realization ratio to identify potential drivers behind a good or bad performance. The results based on the testing pro-jects show among the top projects with a high total realization ratio the important similar factor is a high further realized ratio. In other words, projects that identify risks that occur during the project execution well typically reported a better overall risk identification performance. The second group of projects is those projects with a high new item ratio, whether those risks all happened or not. During the project construction, these projects were active in identifying new potential risks, with an acceptable overall risk realization performance. Finally, projects with low performance in risk identification during project execution reported low overall performance.

Secondly, the initial realization ratio was 0.56 on average. About half of identified risks in the first step did not happen. That is potentially related to the lack of detailed information in the initial phase of the project which makes risk identification more generic rather than project-specific. However, the further realized ratio, which is related to the occurrence of identified risks during the construction process, is higher, on average 0.73. While the overall realization ratio and further realization ratio are higher than the initial realization ratio, the efficiency ratio was 0.43 on average. The potential reason may be that project teams do not take the project risk monitoring and updating seriously. In other words, teams may identify a few risks in the first step and not update the risk registers through project execution. This observation can be found with the new item ratio reported as 0.50 on average. That means 50% of identified risks were considered during the project execution as new risk items. Similarly, the variation is high, reported as between 0.06-0.75.

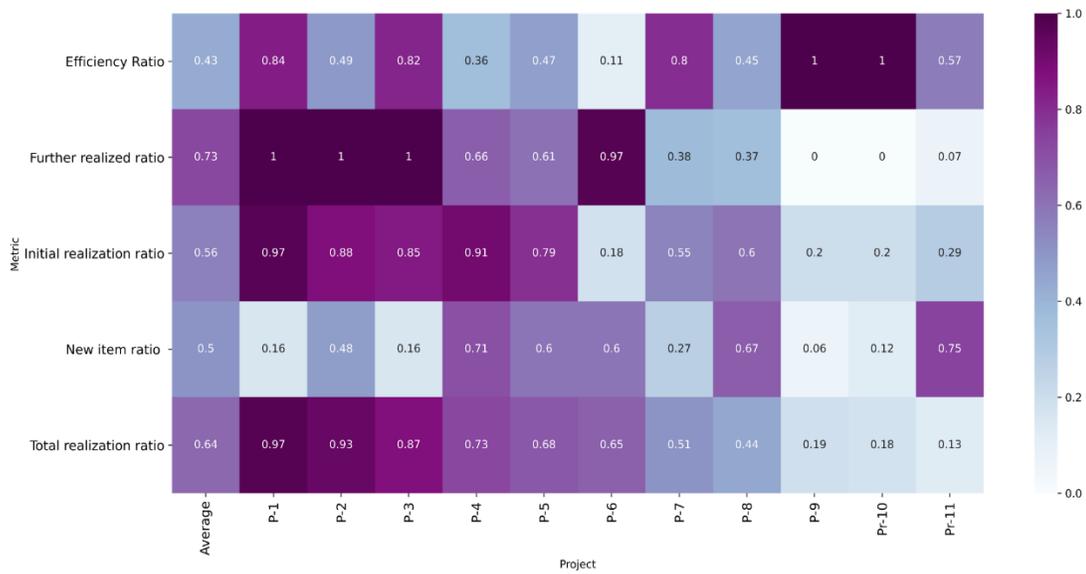

Figure 15. Risk identification performance

4.3.2 Risk Management Styles

Analyzing the risk identification performance metrics suggested categories of potential risk management styles. The study introduces new terms for risk management styles based on the project team performance during the initial and project execution stages. Based on the observed behaviors, generally, project teams can be categorized into two main groups "planners" and "doers". Planner behavior is associated with those teams that mostly try to identify risks in the initial phase of the project and be well prepared for the future. Those teams are not significantly active during project execution to update the risk registers. On the other hand, doer behavior describes project teams that are conducting risk identification actively during project execution. A sign of a doer behavior is found in the new item ratio. Those teams reporting high values for the new item ratio are those who constantly add new risks to the risk register during project execution.

It is important that how the project team's behavior attitude. If a team had planer behavior did they identify a large number of risks realized at the final stage of the project? Or if they were active during project execution how well do they identify risks? Project teams are divided into two major categories, careful and excessive teams. If a team performed well in the initial risk identification it was categorized as "Careful planners." A careful planner team has a high initial realization rate, meaning they identify many risks early on that ultimately occurred. "Excessive planners" are those project teams that identify a large number of risks in the initial stage that did not finally occur, an indication of a low initial realization ratio. Similarly, project teams with careful doer behavior were called "Careful doers." A careful doer team identifies more risks during project execution and those risks finally occurred. While the excessive doer did not perform well in identifying risks during project execution in terms of occurrence with a low further realized ratio. Figure 16 demonstrates the risk management styles considering the proposed metrics, assigned to each project.

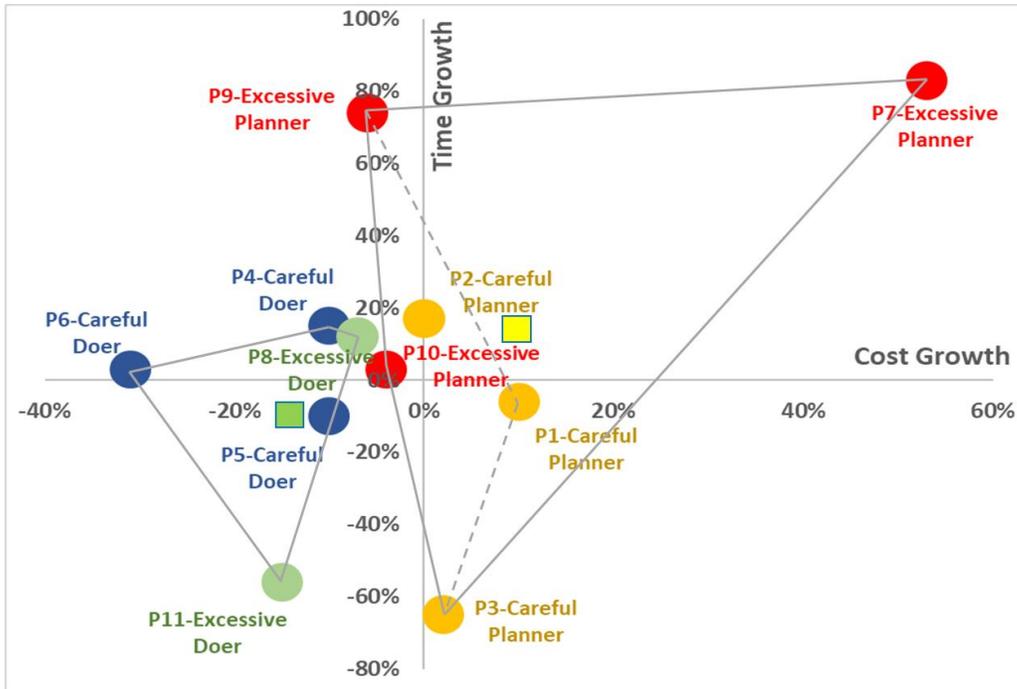

Figure 16. Project delivery performance and risk management style

4.3.3 Project performance vs risk management styles

The purpose of this section is to evaluate the relationship between risk performance metrics, risk management styles, and project delivery performance. Figure 14 provides the project delivery performance metric comprising "Total cost growth," and "Total time growth." Total cost/time growth represents the growth from the engineer's estimate to actual cost/time. In this regard, for example, -0.10 as total cost growth means the project delivered with 10% lower cost than the engineer's estimate. On the other hand, 0.17 as total time growth represents the project delay of 17% compared to the engineer's estimate.

While it is accepted that there are several drivers behind good or bad project delivery, the purpose of this analysis is to evaluate the potential relationship between risk identification style and project delivery performance. According to Figure 16, successful project teams that deliver projects under budget and within schedule are mostly "doer" groups in the current definition. Specifically, careful doers (Projects 4,5, and 6) delivered their projects with good performance metrics. Even projects 11 and 8 which were among the excessive doers delivered their projects well. On the other hand, the more unsuccessful project deliveries (Projects 1, 2, 7, and 9) were planners. This observation indicates careful planning is necessary to deliver a project on time and within the targeted budget but not sufficient. Active risk identification and monitoring behavior during project execution is also required for successful project delivery.

We performed a quantitative analysis to evaluate the statistical significance of the average difference between the performances of the "doer" and "planner" teams. Visual inspection of the data in Figure 16 suggests a distinct difference, but the numbers of data are small. First, the labelled data were clustered into two groups for the doer and planner teams, respectively. The data were normalized into z-scores by dividing by the respective standard deviations to create a common variance. Then Hotelling's $T^2$ test—a generalization of the Student *t* test in one dimension (Rencher 2002)—was applied to test the significance of the difference between the bivariate mean performances of each of the two clusters. This led to the result $T^2 = 27.8$, compared with the *p*=5% limit $T^2_{v=6,p=0.05} = 13.9$. To test the proposition that project P7 is potentially an outlier, that datum was removed and the analysis performed again. Absent P7 the Hotelling statistic became $T^2 = 17.9$, compared with the 5% limit $T^2_{v=5,p=0.05} = 17.3$. Thus, even with this possible outlier removed, the difference between the average performances of the doers and planners remains statistically significant at *p*=5%.

### 4.3.4 Discussion

As a best practice to ensure project success, risk management has been widely implemented in major transportation projects. However, the current risk management practice and most of the previous research studies completed only considered the ex-ante analysis. There is a lack of study to evaluate how the current risk management practice performs in terms of identifying the risks and how its performance link to the project delivery performance. The study introduces a novel framework to track risk item evolving paths that provide an opportunity to evaluate the risk identification performance.

Capitalizing on a data-driven approach this study identifies potential drivers behind a good risk identification practice and a successful project delivery. The main output of this research is that considering risk management as a compliance requirement to complete in the initial phase of a project is a big mistake by project teams. Active and careful risk monitoring during project execution is an essential requirement of successful project delivery. Analyzing the risk management behaviors motivates authors to introduce new terms as project team risk identification styles. Findings revealed that careful planning is necessary for successful project delivery but not enough. A careful doer plan is an essential component of a good risk management practice. Those teams that were active in project execution to identify new risks and monitor the project changes were delivered the project under planned budget and schedule mostly. While inactive teams in terms of risk monitoring delivered projects mostly with huge delay and cost overrun. The main objective of the study is to help project teams be better able to anticipate how such project risks will change as projects move forward and to be better able to forecast changes to the risk register from *ex-ante* to *ex-post* conditions.

## 4.4 Conclusion

The study offers a data-driven framework to evaluate risk identification performance using historical risk data. The framework is informed by automata theory to define risk states and transition functions to track risk life-cycles. Risk states are categorized as registered, happening, and closed. The risk transition functions were categorized as generate, continue, occur, and close. Through tracking individual risk life-cycles, the study introduces new metrics to measure risk identification performance. Performance metrics are designed to figure out the percentage of risks that occurred or were dismissed as the project evolved. These metrics are categorized to measure the risk identification performance in total, initial, and project execution levels.

The authors provided the application of the proposed framework by testing 11 major transportation projects built in the U.S. The results revealed that on average about 64% of identified risks occur through the entire life-cycle of projects. Project teams reported significantly different risk management styles. While some teams try to identify most of the risks in the first step at the initial phase, some teams were significantly active during project execution to identify new risks and update the risk registers.

The study introduced new terms to categorize a project teams' risk style based on their planning and doing behaviors. A *careful planner* style identifies many risks that are realized at the end of the initial stage. A *careful doer* style identifies many risks that happen during project execution. On the other hand, *excessive planners and doers* stand for teams that report low performance in initial realization ratio and further realized metrics, respectively. Finally, the study also examined how doer and planner teams perform differently when it comes to project delivery by applying statistical test. We found that project teams with active and careful doer styles to monitor risks during project execution performed better final cost and time performance that the project delivered compared to the engineer's estimate.

Study limitations include relatively small sample size and limited risk occurrence metrics. Future studies could increase the number of risk registers for comparison and consider the actual consequences of risks in cost and schedule as compared to estimations by the project team. Noted, in the current practice, risk data is limited or not well documented. The study encourages practitioners and researchers for more transparent practice in risk management domain.

# CHAPTER 5: A COMMON RISK BREAKDOWN STRUCTURE AND RISK INTERDEPENDENCIES

The contents of Chapter 5 are accepted to be published by IEEE Transactions on Engineering Management.

**Citation**: Erfani, A., Cui, Q., Baecher, G., Kwak, Y. H. (2023c). Data-driven approach to risk identification in major transportation projects: A common risk breakdown structure, *IEEE Transactions on Engineering Management.*

## 5.1 Abstract

Identifying and evaluating risks is one of the most essential steps in risk management in construction projects. When technical and managerial complexity increases in major transportation projects, this becomes even more important. Currently, project teams are assumed to identify risks mostly based on their experience and expertise. It is a major issue that some state departments of transportation (DOT) project teams lack the risk management experience. This study proposes using a data-driven approach to unify and summarize existing risk documents to create a comprehensive risk breakdown structure (RBS). As a preliminary risk identification framework, a consolidated RBS were developed, using content analysis of public risk reports by various DOTs. Then, comparison was made between the developed RBS with 70 US transportation projects' risk registers. Natural language processing techniques, Bidirectional Encoder Representations from Transformers (BERT), was employed to calculate semantic text similarity to determine what percentage of risks are covered by generic RBS. The results showed that 70 generic risk templates cover almost 81% of the identified risks in the database of 70 major projects which is about 6,000 individual risks. Project parties can use these results to discuss and identify context-specific risks as a starting point. The study also determined the interactions between risk items based on their co-occurrence using historical data. Research findings revealed the importance of considering interdependencies between risks in future studies.

**5.2 Research Design and Data**

Figure 17 illustrates the research flowchart for this study. Data collection, RBS development, RBS testing, and risk interdependencies examination are the four major steps in the development of this study. The following sections provide more details.

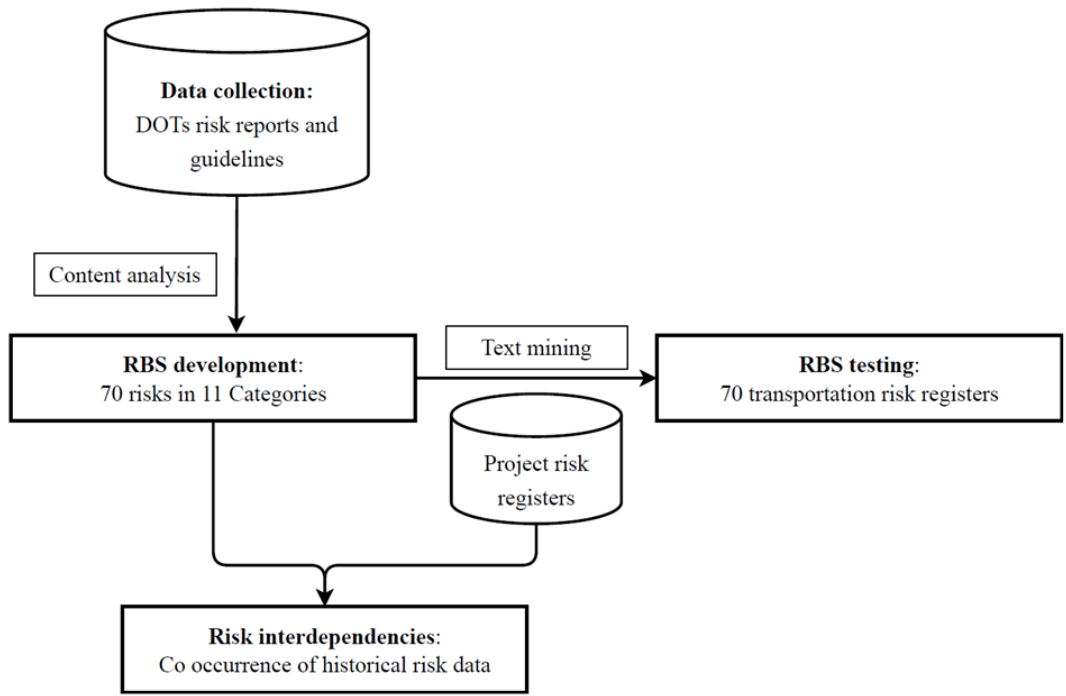

Figure 17. Data-driven risk breakdown structure development outline

5.2.1 Data Collection

A review of DOT websites across the US and other resources was conducted to find out if they have risk management guidelines, reports, or best practices for risk management in construction transportation publicly available. We included risk reports with RBSs among our sample. DOT reports summarize decades of project delivery experience. The final sample includes data from Washington, Texas, Montana, California, Nevada, Michigan, New Jersey, Oregon, and Minnesota, as well as a report from the Federal Highway Administration (FHWA). The raw data, including risk reports and RBS samples, can be found at https://github.com/data-driven-RBS/usingreports.

5.2.2 Risk Breakdown Structure (RBS) Development

DOTs' risk reports group risks into level 1 (risk categories) and level 2 (risk items). Using content analysis, we summarized the risk data into one comprehensive RBS. The content analysis was conducted using a bottom-up approach (Sigmund and Radujković 2014). Risk items were grouped into themes to identify potential categories (level 1) and risk items (level 2) (Rasool et al. 2012). A continuous grouping process similar to (Beardmore and Molenaar 2021) was followed, and iterations continued until: (1) all the categories and items were mutually exclusive and exhaustive, so there could be no unit that falls into two categories or has two points representing it (Krippendorff 2018); (2) each group consisted of a logical quantity so that the grouping was easy to recall (Wicks 2017); (3) ensure diversity of risk factors; and (4) balance the frequency of risks coded to each category (Beardmore and Molenaar 2021, Krippendorff 2018). As soon as all four conditions are met, the RBS is prepared by combining the final categories and risk items.

For example, our first step in developing level 1 is to extract all risk categories and risk items themes within those categories. The initial list includes environmental, design, right-of-way, construction, external, organizational, management, stakeholder, scope, financial and economic, contracting and procurement, railroads, utilities, traffic, market conditions, and structure and geotechnical issues. Our initial grouping process showed that external and stakeholder risks are the same, so we merged them. As there were few risks under scope, railroad, market conditions, and financial and economic, we followed other project solutions and brought those risks to level 2 and merged them with other related categories to balance risk categories. This example demonstrates how content analysis is performed.

### 5.2.3 RBS Testing

This section evaluated the performance of the developed RBS by using the 70 transportation projects (Figure 8). Using Natural Language Processing (NLP) techniques, we identify the risks in testing projects that match the RBS the closest. RBS performance was measured using semantic text similarity calculations. As different project teams use different words and phrases to describe risks, implementing NLP modeling enables us to identify similar risks automatically.

An initial step in NLP studies is text vectorization, which involves converting text into numeric form for further calculations (Ge and Moh 2017, Panahi et al. 2023). Artificial intelligence companies such as Google and Facebook built large pre-trained word embedding vectors. Word embedding models show each particular word in the corpus while capturing the semantic meaning behind words. One of the most widely used vector models in the literature is Word2Vec introduced by Google's research team (Mikolov et al. 2013). The Word2vec model was trained with millions of words from Google news, which were converted into 300-dimension vectors based on their meaning.

The deployment of deep learning models with transformer architecture to generate contextualized word embeddings has led to a tremendous improvement in NLP models in recent years. BERT (Bidirectional Encoder Representations from Transformers) is a state-of-the-art pre-trained language model developed by Google. BERT has the advantage of generating contextualized embeddings when compared to word2vec. Using word2vec, each word is represented by a fixed vector in a high-dimensional space, which does not capture its context-specific meaning. By contrast, BERT generates word embeddings based on the context of words in sentences or paragraphs using a transformer architecture. Thus, BERT can accurately represent words in various contexts by capturing the nuances of language (Kenton et al. 2019; Reimers and Gurevych 2019). BERT has been applied to tasks such as information retrieval, sentiment analysis, and document classification in construction-related studies. Using BERT, Fang et al. (2020) classified texts related to construction safety in a recent study. BERT was used by Moon et al. (2022) for automatic risk identification in contracts.

SBERT (Sentence-BERT) is a variant of BERT (Bidirectional Encoder Representations from Transformers) that is specifically designed for computing sentence embeddings and measuring semantic similarity between pairs of sentences (Reimers and Gurevych 2019). First, the input sentences are preprocessed to add special tokens that indicate the beginning and end of the sentences, along with a separator token that distinguishes them. Using a large corpus of text data, the pre-trained BERT model is fine-tuned for similarity or paraphrasing tasks. As soon as the model is fine-tuned, it can be used for generating embeddings of any input sentence by passing the sentence through the model and extracting the embedding from the output. A pooling layer is applied to the BERT model to obtain sentence-level embedding. The embedding vector of a text with 768 dimensions is what we get after the pooling layer. Using pairwise distances or cosine similarity, these embeddings can be compared. Cosine similarity is a general measure of assessing the similarity of two vectors by calculating the cosine of their angles. A calculated example showing the semantic similarity calculation process between two risk items is shown in Figure 18. Cosine similarity calculations facilitate the matching of similar risks within RBS and testing risk registers that measure RBS performance.

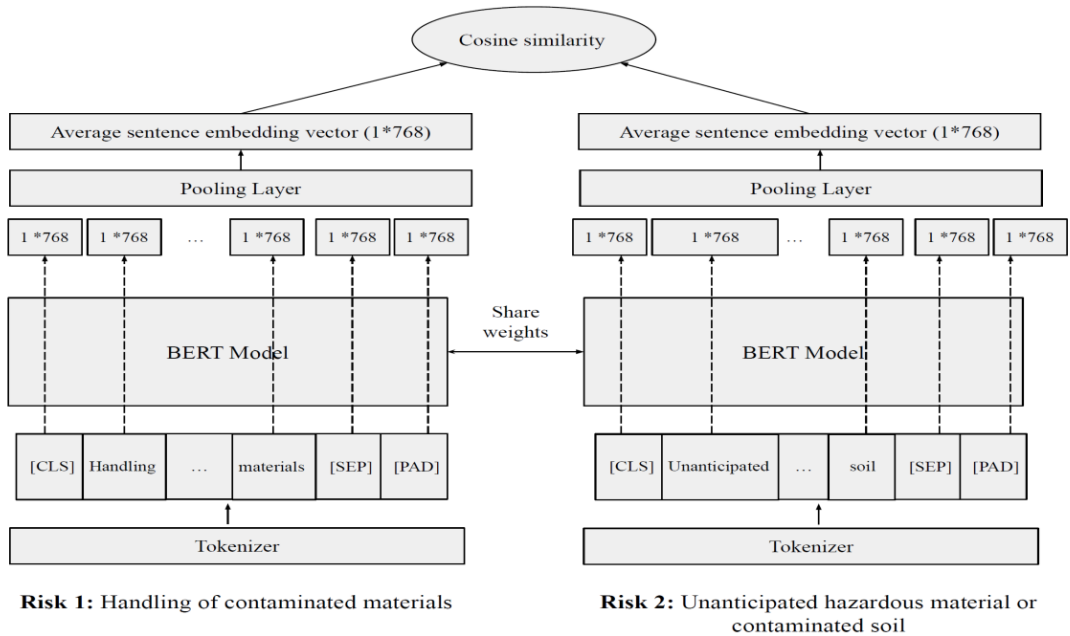

Figure 18. Semantic similarity calculation example using BERT

Ideally, the similarity index should equal 1, meaning the same risk item repeats exactly the same. The threshold for text similarity matching using BERT depends on the specific use case and the level of similarity required. A threshold of 60% cosine similarity is often used by BERT to identify sentences that are similar or paraphrased (Reimers, N. and Gurevych 2019; Kasnesis et al. 2021; Zhang et al. 2019). We found that 60-70% similarity is an appropriate threshold for matching risk across different texts. It should be noted that our text similarity matching approach has potential limitations in terms of mislabeling, especially if the similarity is between 50-70% for both risks incorrectly considered as cover and the reverse.

5.2.4 Risk Co-occurrence Examination

With a comprehensive RBS and historical data, we can measure the co-occurrence of risks across 70 major transportation projects. Based on the developed RBS in step B and the result of matching risks to closet risks in step C, we get an output of each risk in RBS occur on which projects. Our next step was to develop a Python script that would perform a pairwise comparison of all the risks in the RBS to determine how many co-occurrences there were between them.

**5.3 Result and Discussions**

5.3.1 Risk Breakdown Structure

A total of 70 risk items and 11 risk categories are included in the final RBS. A visual representation of the risk categories in level 1 is shown in Figure 19. Based on the DOT's report, we discuss the detail of risks in each category and their definitions.

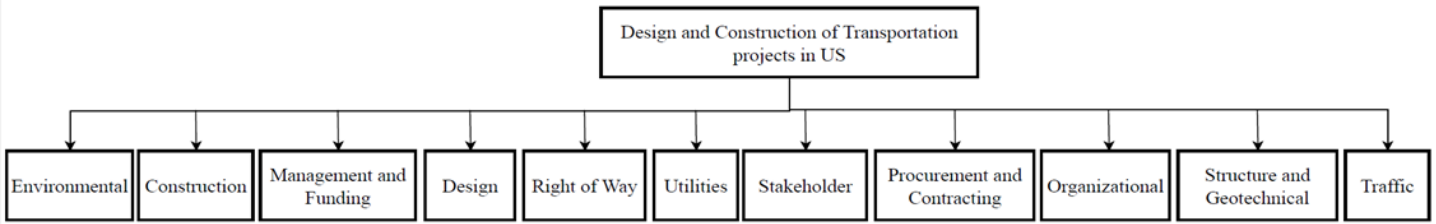

Figure 19. RBS level 1 for transportation projects in US

*Environmental*

The 'Environmental' risk category includes risks associated with processes and permits related to the National Environmental Policy Act and State Environmental Policy Act, hazardous material (Washington DOT 2018). This category includes following major risk items: "Environmental permitting and requirements", "National Environmental Policy Act Review (NEPA) process and documentation", "Hazardous Materials", "Wetlands and endangered species", "Archaeological and cultural sites", "Environmental regulation change", "Additional environmental analysis required", "Water Quality", "Noise mitigation", and "Unidentified contaminated soils".

*Construction*

The 'Construction' risk category includes risks associated with construction phase of the project with emphasize on schedule related, safety and earth work issues (Dicks and Molenaar 2022; Washington DOT 2018). Many of the identified risks in the Construction category overlapped with those in other categories but could be classified as construction risks due to when they occurred (Dicks and Molenaar 2022). This category includes following major risk items: "Contractor access", "Different site and subsurface condition", "Construction Safety", "Schedule uncertainty", "Coordination with adjacent projects", "Work windows", "Material and resources availability", "Construction incorporates new or unproven technology", "Contractor and subcontractor performance", "Weather related issues", "Buried man-made objects", and "Construction quality assurance and control issues".

*Management and funding*

The 'Management and funding' risk category includes risk associated with managerial aspect, decision making, workforce limitation, economic and financial management (Dicks and Molenaar 2022; Washington DOT 2018). This category includes following major risk items: "Delayed decision making", "Project purpose change", "Cash flow restrictions", "Labor disruptions", "Force majeure", "Economic change and availability of funding", and "Political or policy changes".

*Design*

In the 'Design' risk category, there are risks associated with the design process, approval, errors, changes, and deviations. There were multiple possible scenarios during the planning phase of the risk workshop, which is usually cited as a reason to include this risk (Dicks and Molenaar 2022; Washington DOT 2018). This category includes following major risk items: "Design changes", "Design requirement", "Design incomplete", "Delay in design approval", "Design exceptions", and "Aesthetic issues".

**Right of Way**

The 'Right of way" risk category includes associated with land acquisition, limited access, and right of way planning. While there may not be as many separate risks identified here, they are undoubtedly one of the costliest risks. As these projects typically require dozens of land acquisitions and the possibility of any of them requiring settlement in court could delay the project timeline (Dicks and Molenaar 2022; Washington DOT 2018). This category includes following major risk items: "Right of way acquisition issues", "Right of way cost uncertainty", "Additional Right of way is required", "Right of way plan", "Railroad and right of way entry", and "Right of way relocation".

*Utilities*

The 'Utilities' risk category includes risks associated with utility relocations that typically require coordinating with local municipalities and utility companies as well as discovery of unexpected utilities that may need to be moved (Dicks and Molenaar 2022; Washington DOT 2018). This category includes following major risk items: "Utility coordination", "Utility requirement", "Utilities conflicts", "Utility funding may be inadequate", and "Utility relocation".

*Stakeholder*

In the 'Stakeholder' risk category, there are risks associated with third parties and public involvement (Dicks and Molenaar 2022; Washington DOT 2018). This category includes following major risk items: "Public involvement", "Additional Scope for third parties", "New stakeholders emerge and demand new work", "stakeholders request late changes", "Objection from local communities and agencies", and "Communication with stakeholders".

*Procurement and Contracting*

In the 'Procurement and Contracting' risk category, there are risks associated with delivery of the project, the execution of the contract, and market conditions that may affect the level of competition before awarding the contract (Dicks and Molenaar 2022; Washington DOT 2018). This category includes following major risk items: "Change in delivery method", "Market condition", "Contract language and legal issues", "Change order and claim", and "Delays in procurement".

*Organizational*

'Organizational' risks are related to policies, guidelines, procedures, and cultures within organizations (Washington DOT 2018). An organization's ability to navigate risk-taking and risk-reduction activities is essential for their success in delivering projects (Crispim et al. 2018). This category includes following major risk items: "Change in leadership", "Organizational resources", "Project dependencies", and "Organizational policy and prioritization".

*Structure and Geotechnical*

Risks associated with foundation design, excavation and geotechnical activities, and soil conditions fall into the 'structure and Geotechnical' category (Dicks and Molenaar 2022; Washington DOT 2018). This category includes following major risk items: "Soil and geotechnical conditions", "Construction excavation", "Pile driving noise and vibration", and "Structural foundation design".

*Traffic*

The 'Traffic' risk category relates to future highway users, traffic growth and revenue specifically for public-private partnerships (PPP) projects (Montana DOT 2016). This category includes following major risk items: "Traffic growth", "Toll related issues", "Bicyclist and pedestrian recommendations may not be supported", "Unanticipated Mobility and/or traffic delays", and "Land use changes".

Table 21 provides an overview of the complete RBS, including the frequency of risk reports in risk documents. Frequency refers to the number of occurrences of risk across 10 collected risk reports.

Table 21. Risk breakdown structure for major transportation projects in U.S.

| Level 1 | Level 2 | Frequency (across 10 reports) |
|---|---|---|
| Environmental | Environmental permitting and requirements | 10 |
| | National Environmental Policy Act Review (NEPA) process and documentation | 8 |
| | Hazardous materials | 6 |
| | Wetlands and endangered species | 5 |
| | Archaeological and cultural sites | 4 |
| | Environmental regulation change | 5 |
| | Additional environmental analysis required | 7 |
| | Water quality | 4 |
| | Noise mitigation | 5 |
| | Unidentified contaminated soils | 4 |
| Construction | Contractor access | 4 |
| | Different site and subsurface condition | 7 |
| | Construction safety | 5 |
| | Schedule uncertainty | 5 |
| | Coordination with adjacent projects | 3 |
| | Work windows | 4 |
| | Material and resources availability | 8 |
| | Construction incorporates new or unproven technology | 3 |
| | Contractor and subcontractor performance | 7 |
| | Weather related issues | 3 |
| | Buried man-made objects | 3 |

| | Construction quality assurance and control issues | 3 |
|---|---|---|
| Management and Funding | Delayed decision making | 3 |
| | Project purpose/scope change | 6 |
| | Cash flow restrictions | 3 |
| | Labor disruptions | 6 |
| | Force majeure | 3 |
| | Economic change and availability of funding | 6 |
| | Political or policy changes | 8 |
| Design | Design changes | 7 |
| | Design requirement | 3 |
| | Design incomplete | 2 |
| | Delay in design approval | 3 |
| | Design exceptions | 3 |
| | Aesthetic issues | 4 |
| Right of Way | Right of way acquisition issues | 9 |
| | Right of way cost uncertainty | 5 |
| | Additional Right of way is required | 5 |
| | Right of way plan | 4 |
| | Railroad and right of way entry | 8 |
| | Right of way relocation | 3 |
| Utilities | Utility coordination | 6 |
| | Utility requirement | 2 |
| | Utilities conflicts | 2 |
| | Utility funding may be inadequate | 2 |
| | Utility relocation | 6 |
| Stakeholder | Public involvement | 8 |
| | Additional Scope for third parties | 5 |
| | New stakeholders emerge and demand new work | 4 |
| | Stakeholders request late changes | 5 |
| | Objection from local communities and agencies | 7 |
| | Communication with stakeholders | 6 |
| Procurement and contracting | Change in delivery method | 3 |
| | Market condition | 9 |
| | Contract language and legal issues | 9 |
| | Change order and claim | 5 |
| | Delays in procurement | 4 |
| Organizational | Change in leadership | 5 |
| | Organizational resources | 4 |
| | Project dependencies | 3 |
| | Organizational policy and prioritization | 4 |
| Structure and Geotechnical | Soil and geotechnical conditions | 3 |
| | Construction excavation | 3 |
| | Pile driving noise and vibration | 3 |
| | Structural foundation design | 7 |
| | Traffic growth | 9 |
| | Toll related issues | 4 |

| Traffic | Bicyclist and pedestrian recommendations may not be supported | 4 |
| | Unanticipated Mobility and/or traffic delays | 3 |
| | Land use changes | 3 |

5.3.2 Risk Breakdown Structure Testing

We compared 70 transportation project risk registers with RBS to see how well it performs as a starting point for identifying common risks and giving the project team more time to focus on project-specific risks. Using semantic text similarity calculation, each risk in the risk register is matched with the highest similar risk in the RBS. As explained in the methodology section, when project teams use their own language and add more project-specific terms, the two risks are similar, but they share a lower text similarity. For example, the risk "Utility conflicts on arterials near Sheridan street" is similar to "Utility conflicts", or the risk "Worker injury during construction or OSHA violation" is similar to "Construction safety" but has more project-related context. According to our observation, 60% text semantic similarity is the threshold for meaningful matching of a generic risk from RBS with a project specific context risk. In Table 22, a variety of risk-matching examples are presented for different levels of semantic similarity.

Risks matched with a risk with more than 60% textual semantic similarity can be divided by all risks to calculate the RBS coverage of the risk register database. The findings show that RBS covers almost 81% of the risks in the database. Figure 20 illustrates in detail the percentage of database risks that match RBS at different semantic similarity levels. As part of our analysis of RBS's performance, we evaluated whether high-impact costs and schedule risks were covered by RBS or not. Based on the project size, cost and schedule impacts of risks are normalized. Covered risks by RBS have an average cost impact of 6.08 Million $, while not covered projects risks have an average cost impact of 1.53 Million $. The schedule impact of covered risks is on average 1.71 months, whereas the impact of not covered risks is 0.84 months. According to findings, data-driven RBS covers 81% of project risks, as well as impactful risks in terms of cost and schedule.

Table 22. Risk matching examples using BERT

| Risk (1): Project risk | Similarity | Risk (2): RBS risk |
|---|---|---|
| Right of way acquisition issues | 1 | Right of way acquisition issues |
| Market condition | 1 | Market condition |
| Excavation operation | 0.95 | Construction excavation |
| Third Party Utility Relocation | 0.92 | Utility Relocation |
| Unanticipated cultural or archaeological findings | 0.84 | Archaeological and cultural sites |
| Subsurface Conditions | 0.81 | Different site and subsurface condition |
| Traffic and Revenue | 0.79 | Traffic growth |
| Poor Soil Permeability Rates | 0.76 | Soil and geotechnical conditions |
| Delay to record of decision - Cost of Schedule Recovery | 0.72 | Delayed decision making |
| No areas for Contractor use identified and cleared. | 0.69 | Contractor access |
| Change in tolling equipment technology | 0.64 | Toll-related issues |
| Termination of DB contract | 0.59 | Contract language and legal issues |
| Opportunity to reduce shoulders width | 0.52 | Additional right of way is required |
| Security requirements | 0.44 | Design requirement |

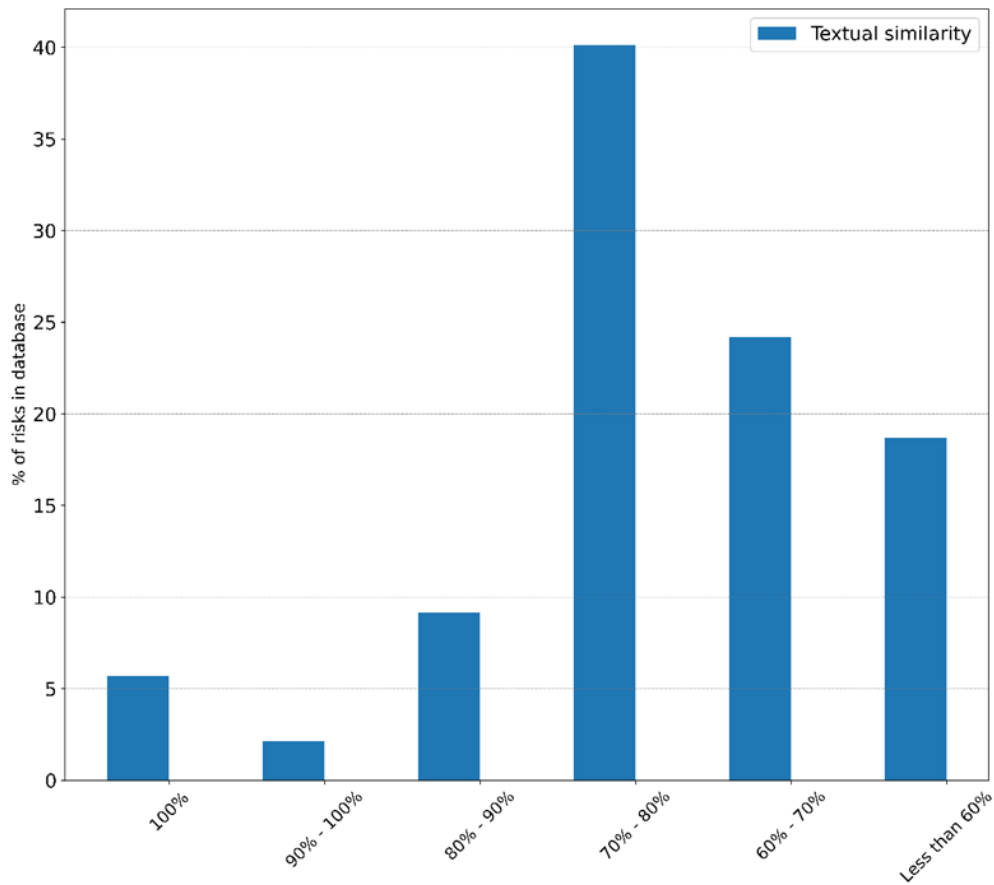

Figure 20. RBS coverage of risk database

We also evaluated the coverage of risk categories and groups among RBS's covered risks. Construction, environmental, and right-of-way risks are among the most prevalent covered risk categories, as shown in Table 23. In addition, the risk categories in actual risk registers allow us to compare our risk categories based on RBS's level 1 risk groups with the risk classes defined by the project team. Most project teams also use the same structure and language to cover risk groups. The purpose of developed RBS was to include a comprehensive list of risk groups. Some scenarios differ from our developed RBS's level 1, including using different terms for risk groups, moving risk items between levels 2 and 1 as classes, and grouping risk classes. For example, projects might refer to stakeholder risk groups under different terms such as partnerships, externals, or third parties. Secondly, project teams sometimes move risk items from level 2 to a risk category like railroads or legal issues, depending on the project. Finally, sometimes there is a difference in how the project team combines some risk classes, such as management and planning, design and construction, and management and design. With these few different scenarios, in over 87% of cases, we validated our RBS with a detailed matching of the RBS at level 1 and the project testing risk classes.

Table 23. Level 1 of RBS risk coverage

| Risk Category | Coverage (%) |
|---|---|
| Construction | 23.38 |
| Environmental | 14.70 |
| Right of way | 11.33 |
| Management and funding | 8.84 |
| Design | 7.85 |
| Utilities | 7.42 |
| Structure and geotechnical | 7.16 |
| Procurement and contracting | 5.66 |
| Traffic | 5.20 |
| Organizational | 4.48 |
| Stakeholder | 3.98 |

5.3.3 Risk Co-occurrence Result

By matching the risk database with RBS, it is possible to evaluate how risks co-occurred in the past. From 70 risk registers, we looked at the number of times risks co-occurred; Table 24 gives details on the top and bottom 10 co-occurring risks. Considering risk interaction in risk assessment and preventing risk interdependency in risk treatment is important due to the vast difference in co-occurrence rate. According to table III, a higher co-occurrence rate indicates that the risks occur more frequently and are also more dependent on each other in terms of occurrence. The results also show that risk type and nature play a significant role in co-occurrence rates. There is a higher co-occurrence rate for technical and engineering risks, while the co-occurrence rate for managerial risks is lower. The higher the rate of co-occurrence, the more likely it is that there is a correlation between risks, but to model interdependencies, logical relationships between risks need to be evaluated. Analyzing co-occurrence rates and logical relationships ultimately assists Monte-Carlo risk simulation by considering risk interdependency. In our study, risk co-occurrence rates were determined using historical data, significant differences were demonstrated, and possible directions were identified for assessing risk interdependencies in future study.

Table 24. Risk co-occurrence out of 70 risk registers

| Risk (1) | Risk (2) | Rate |
|---|---|---|
| Right of way plan | Utility relocation | 40 |
| Delay in procurement | Utility relocation | 36 |
| Contractor Access | Utility relocation | 36 |
| Right of way plan | Delay in procurement | 34 |
| Design changes | Utility relocation | 33 |
| Contractor Access | Right of way plan | 33 |
| Contractor Access | Different site and subsurface condition | 32 |
| Contractor Access | Delay in procurement | 32 |
| Hazardous Materials | Utility relocation | 32 |
| Right of way acquisition issues | Utility relocation | 31 |
| Design changes | Right of way plan | 31 |
| Contractor Access | Hazardous Materials | 30 |
| Contractor Access | Design changes | 30 |
| Contractor Access | Construction Excavation | 30 |
| Hazardous Materials | Right of way plan | 30 |
| Risk (1) | Risk (2) | Rate |
| Change order | Unidentified contaminated soils | 1 |
| Change order | Organizational policy and prioritization | 1 |

| New stakeholders emerge and demand new work | Construction safety | 1 |
| Delayed decision making | Organizational policy and prioritization | 1 |
| Different site and subsurface condition | Organizational policy and prioritization | 1 |
| Cash flow restriction | Objection from local communities and agencies | 1 |
| Environmental regulation change | Unidentified contaminated soils | 1 |
| Bicyclist and pedestrian recommendations may not be supported | Objection from local communities and agencies | 0 |
| Land use changes | Organizational policy and prioritization | 0 |
| Bicyclist and pedestrian recommendations may not be supported | Organizational policy and prioritization | 0 |
| New stakeholders emerge and demand new work | Organizational policy and prioritization | 0 |
| Organizational resources | Additional Scope for third parties | 0 |
| Change in delivery method | Organizational policy and prioritization | 0 |
| Organizational resources | Change order and claim | 0 |
| Aesthetic issues | Objection from local communities and agencies | 0 |

## 5.4. Discussion

The current risk identification practices heavily rely on input from subject matter experts. Project teams perform risk analysis using their expertise and experience. A major challenge of current risk management practices is the lack of organized historical data, the failure to learn from past experiences, and the reliance on experts only. Therefore, data-driven objective approaches can be applied for risk detection with the advancement of technology and the availability of historical data. The purpose of this study was to develop a data-driven RBS by using risk management guidelines and reports from several DOTs. Using a comprehensive RBS, findings showed that lessons learned in the previous project could be used to identify more than 80% of risk items in 70 testing transportation projects. In addition to covering frequent risks, RBS also covers high cost/schedule impact risks. Project risks covered by RBS report a higher cost and schedule impact on average than those not covered. Utilizing lessons learned from past projects, data-driven risk modeling provides project teams with an advisory tool to complete their expert evaluation. Furthermore, data-driven RBS provides a mechanism that can be compared to actual risk registers to identify potential risk networks and co-occurrences, which ultimately helps in modeling risk interdependencies.

When comparing the performance of developed RBS based on different project characteristics, project delivery method and project type have greater influences on risk detection than project size. The comprehensive RBS covered 83.6% of risks in Design-Build-Build (DBB) projects, and 80.4% of risks in Design-Build-Build (DB) projects, but only 75.4% of risks in PPP projects. This observation concurs with our previous study in terms of the higher risk uniqueness of PPP projects compared to traditional delivery methods projects [40]. As for project types, highway projects perform the best with 84.4% and roadway projects with 82.1%, while interchange projects, bridges and tunnels only have 75.2% and 72.7% risks covered by RBS, respectively. The reason for this observation is that most source reports relate to highway and roadway projects, resulting in less coverage of specific context risks in bridge and tunnel projects. By looking at the project size, RBS covers 82.3%, 81.9%, and 78.5% of risks in projects under 500 M$, over 1 B$, and 500 M$- 1 B$. No matter how large or small a project is, RBS provides consistent risk coverage on average.

When compared to actual risk registers, "Design changes", "Utility relocations", "Additional Right of Way is required", "Delays in procurement", and "Hazardous materials" are among the most frequent covered risks in developed RBS. With a few exceptions, considering the project characteristics resulted in a similar pattern of high frequent risks. Bridges and tunnels projects, for example, place "construction access" at the top of the list, while PPP projects place more emphasis on "economic change and funding availability". In order to identify some possible trends in uncovered risks, we examine them in detail. In the first instance, adding a lot of context to a risk item may cause it to be uncovered by the similarity calculation while the generic example is included in the template. As an example, while "Toll related issues" is listed in RBS, the risk "Segment tolling vs. trip building Tolling strategy may change the signage requirements for the projects" is labeled uncovered. Among other examples, one solution is to make sub-lists of context-specific risks under each generic risk in RBS. A number of other risks uncovered, such as "Awareness campaign", "Removal of snow and ice due to shading", "Pleasing everyone", "Inadequate lane closures notice to road users", "Culvert replacement", "Air quality", "Union jurisdiction/labor agreements," and several others, could be listed as a group of other risks to provide the users with some less frequent risks as well.

The risks listed in the RBS can be used by any transportation project team conducting a risk analysis. Two possible uses of the RBS might be considered in discussions regarding its implementation: first, to stimulate a discussion about the themes and identify project-specific risks associated with them; second, to serve as a "back pocket" checklist once the initial round of risk identification is completed to ensure all major items are covered. Further, historical risk co-occurrence data reveals the importance of modeling risk interactions and interdependencies in risk management. Risk type plays a significant role in the co-occurrence of risks, particularly technical and engineering risks reporting a higher co-occurrence rate than managerial risks. Additionally, by incorporating a network view into risk assessment, the current hierarchy view of risk management can be improved. To update prior information on causal effect relationships between risks, future studies need to utilize historical evidence.

## 5.5. Conclusion

The inherent characteristics of major transport projects make them highly risky and uncertain. Consequently, cost overruns and delays are major concerns for transportation agencies when implementing these projects. To manage projects effectively, risk management practices seek to detect these challenges, evaluate them, and propose appropriate responses. There are numerous studies that provide different tools and techniques to complete the risk identification process, but industry practices are still based on experience and subject matter experts' opinions. A comprehensive RBS is developed in this study in order to develop a data-driven approach to risk detection. Using natural language processing, we evaluated RBS's performance in detecting early risks in 70 transportation projects. Nearly 81% of risks in the database were covered by the developed RBS. Based on the findings, it is evident that common risks occupy a large portion of the risk register and can be adapted from similar past project experiences using objective data to allow the project team more time to develop project context risks. Moreover, the risk interdependency examination of RBS using historical risk registers illustrates the importance of risk interaction considerations when treating risks.

By using historical real-risk data and incorporating a data-driven RBS, this study contributes to the body of knowledge concerning risk management in transportation projects. One limitation of the study is the use of general word embedding models in semantic similarity calculation. Future research could focus on developing word corpora based on construction context. Furthermore, the RBS developed in this study only include information regarding risk occurrence and do not address the consequences and mitigation strategies. Data-driven approaches could be used in future studies to summarize lessons learned from historical risk mitigation strategies and risk assessment results. As well as this, the availability of relevant risk reports from the DOTs further limited this study. Finally, identifying risks using data-driven methods would be more effective if a larger pool of data were available.

# CHAPTER 6: CONCLUSION

## 6.1. Major findings and novelty

In most cases, risk management of major transportation projects involves expert opinion through risk workshops. This study introduced a novel risk management strategy based on data and artificial intelligence techniques. Despite the fact that each project is unique, a comparison of risk similarity reveals that historical projects have faced a great deal of the same risks. A higher risk similarity is observed among major transportation projects that are of the same type and come from the same area. Therefore, a predictive risk model can be developed to advise project teams about potential risk items and their impacts considering project characteristics such as type, delivery method, location, and size. Findings illustrated that project type and location are major drivers behind risk similarity among major transportation projects. The predictive risk model has been tested on several projects with more than 65% recall rate. In other words, instead of starting from scratch, a team could use the predictive risk model to start with over half of their risks from similar past projects.

Additionally, despite decades of research on risk identification, there have been no empirical studies demonstrating the performance of risk identification and its relationship with project delivery performance. Using automata theory, this study proposed a new framework for tracking individual risk item lifecycles in order to evaluate risk identification performance. The study found that project teams identified 45% of risks out of total risks during the implementation phase when new information about the project became available. Teams with doer behavior, i.e., actively updating risk registers and identifying risks, performed better in delivering the project. Therefore, better risk identification directly contributes to better project delivery.

Lastly, the study investigates whether all historical risk records can be summarized into a comprehensive risk breakdown structure (RBS). Based on content analysis and available risk reports developed by several departments of transportation (DOT), a comprehensive RBS has been developed to include the most frequent risk categories and risk items. There are 70 generic risk items and 11 generic risk categories in the comprehensive RBS, which covers almost 80% of the risks in the database.

**6.2. Managerial implication**

Furthermore, in addition to the study's contribution to theories in risk management by defining data-driven risk identification and risk management performance measurement frameworks, the study also has several immediate implications for practice. To begin with, the RBS and predictive risk model can be used as advisory tools to help project teams identify risks based on similar past projects. Predictive risk models can be extended to web-based tools via a current version of the model developed in Excel.

Secondly, the developed framework based on automata theory can be applied in practice for measuring risk identification performance. The project team can use risk life-cycle modeling to assess their risk identification performance, model their project risk management style, and evaluate the impact of risk management on project delivery. Finally, data driven risk management might work as a complementary tool to help all stakeholders. This may be used by the sponsor of the project to verify the completeness of the risk registers, while it may be used in two ways by the project team: first, to stimulate a discussion and identify project-specific risks associated with the themes; second, to make sure all major items are covered after the initial round of risk identification has been completed.

**6.3. Future research directions**

Results of this study demonstrate the potential application of historical data and artificial intelligence techniques to infrastructure management. There is potential for extending this concept to risk allocation, risk response, and risk documentation in contracts. An important area of research focuses on the automatic detection of risks in a contract and the way they are allocated. Therefore, natural language processing techniques can streamline contract risk reviews. There is also the potential to make a roadmap for a cloud-based shared data-driven risk model that hosts a large dataset of risks and different stakeholders will be able to add new data to improve algorithms (Figure 21).

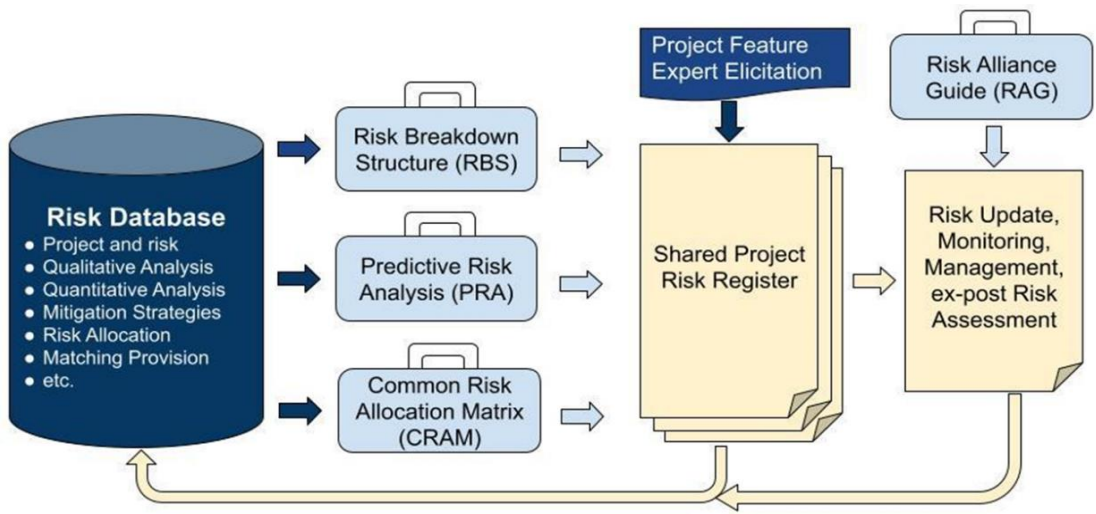

Figure 21. Roadmap of AI-based Risk Management

# Glossary

| Cosine Similarity | A popular method used to measure the similarity between two documents by calculating the cosine value. |
|---|---|
| Deep Learning | A machine learning technique that is often built on multiple layers of artificial neural networks. |
| F1-score | A combination of Precision and Recall, which is used to measure the classification performance on each class. |
| FastText | A word embedding technique released by Facebook's AI Research lab in 2017, which uses the Continuous Bag of Words (CBOW) model to configure the word representation and a hierarchical classifier and N-gram features to improve training efficiency |
| Global Vector (GloVe) | A word embedding technique developed at Stanford University in 2014, in which the resulting representations showcase the structures of the word vector space trained on the co-occurrence probabilities of words. |
| Recall | The fraction of true positives over all the cases that actually are positive |
| TF-IDF | A statistical measure assigned to each word in a document (e.g., a sentence, or a tweet in this study), which can be used to estimate the importance of a word appears in a document given a corpus of documents. |
| Text Vectorization | A process of converting the text into numerical representation (e.g., a vector or a matrix of real numbers). |
| Word Embedding | A word representation technique in NLP in which words or phrases are mapped into dimensional vectors of real numbers. |
| Risk Identification | The mechanism of determining which risk items may affect the project |
| Finite state automation | Finite State Automaton (FSA) is a mathematical model used to describe and represent systems that have a finite number of states and transition between those states based on input from a defined set of symbols. |
| Word2vec | Word2Vec is a word embedding model developed by Google that aims to learn continuous vector representations, or embeddings, for words from large amounts of text data. Word embeddings are dense vector representations that capture the semantic meaning and syntactic relationships between words, making them useful for a wide range of natural language processing (NLP) tasks. |

| BERT | BERT is a natural language processing (NLP) model developed by Google in 2018. BERT is a type of transformer-based model, which is a neural network architecture that is particularly effective at processing sequential data, such as text. Unlike traditional language models that process text in one direction (either left-to-right or right-to-left), BERT is designed to capture contextual information from both the left and right context of a word. |
|---|---|
| Text classification | A machine learning technique that classifies a set of textual data into targeting classes. |

# Bibliography

Abdelgawad, M. and Fayek, A.R., (2010). Risk management in the construction industry using combined fuzzy FMEA and fuzzy AHP. Journal of Construction Engineering and Management, 136(9), pp.1028-1036. https://doi.org/10.1061/(ASCE)CO.1943-7862.0000210

Afzal, F., Yunfei, S., Nazir, M. and Bhatti, S.M., (2021), A review of artificial intelligence-based risk assessment methods for capturing complexity-risk interdependencies: Cost overrun in construction projects, International Journal of Managing Projects in Business, Vol. 14 No. 2, pp. 300-328. https://doi.org/10.1108/IJMPB-02-2019-0047

Al-Bahar, J. F., & Crandall, K. C. (1990). Systematic risk management approach for construction projects. Journal of construction engineering and management, 116(3), 533-546. https://doi.org/10.1061/(ASCE)0733-9364(1990)116:3(533)

American Society of Civil Engineering (ASCE)., (2021). A comprehensive assessment of America's infrastructure. https://www.infrastructurereportcard.org/wp-content/uploads/2020/12/2021-IRC-Executive-Summary.pdf

Anari, B., Ahmadi, M. R., gobaei Arani, M., & Anari, Z. (2013). Optimizing Risk Management Using Learning Automata. International Journal of Computer Science Issues (IJCSI), 10(3), 313.

Baiardi, F., Martinelli, F., Ricci, L., Telmon, C., & Pontecorvo, L. B. (2008). Constrained automata: a formal tool for risk assessment and mitigation. Journal of Information Assurance and Security, 3, 304-312.

Beardmore, D. C., & Molenaar, K. R. (2021). Roadway design and construction in infrastructure limited contexts: a risk breakdown structure. International Journal of Construction Management, 1-9. https://doi.org/10.1080/15623599.2021.1999768

Bhattacharya, C., & Ray, A. (2022). Thresholdless Classification of chaotic dynamics and combustion instability via probabilistic finite state automata. Mechanical Systems and Signal Processing, 164, 108213. https://doi.org/10.1016/j.ymssp.2021.108213

Bojanowski, P., Grave, E., Joulin, A. and Mikolov, T., (2017). Enriching word vectors with subword information. Transactions of the Association for Computational Linguistics, 5, pp.135-146. https://doi.org/10.1162/tacl_a_00051

Chen, Q., & Mynett, A. E. (2003). Effects of cell size and configuration in cellular automata based prey–predator modelling. Simulation Modelling Practice and Theory, 11(7-8), 609-625. https://doi.org/10.1016/j.simpat.2003.08.006

Creedy, G.D., Skitmore, M. and Wong, J.K., (2010). Evaluation of risk factors leading to cost overrun in delivery of highway construction projects. Journal of

Construction Engineering and Management, 136(5), pp.528-537. https://doi.org/10.1061/(ASCE)CO.1943-7862.0000160

Crispim, J., Silva, L. H., & Rego, N. (2018). Project risk management practices: the organizational maturity influence. International journal of managing projects in business, 12(1), 187-210. https://doi.org/10.1108/IJMPB-10-2017-0122

Cui, Q., & Erfani, A. (2021). Automatic detection of construction risks. In ECPPM 2021–eWork and eBusiness in Architecture, Engineering and Construction (pp. 184-189). CRC Press.

Curtis, J. A., & FHWA International Technology Scanning Program. (2012). Transportation risk management: International practices for program development and project delivery. US Department of Transportation, Federal Highway Administration, Office of International Programs.

De Caso, G., Braberman, V., Garbervetsky, D., & Uchitel, S. (2009). Validation of contracts using enabledness preserving finite state abstractions. In 2009 IEEE 31st International Conference on Software Engineering (pp. 452-462). IEEE. https://doi.org/10.1109/ICSE.2009.5070544

De Caso, G., Braberman, V., Garbervetsky, D., & Uchitel, S. (2010). Automated abstractions for contract validation. IEEE Transactions on Software Engineering, 38(1), 141-162. https://doi.org/10.1109/TSE.2010.98

De Winter, J. C. (2013). Using the Student's t-test with extremely small sample sizes. Practical Assessment, Research, and Evaluation, 18(1), 10.

Di Giuda, G.M., Locatelli, M., Schievano, M., Pellegrini, L., Pattini, G., Giana, P.E. and Seghezzi, E., (2020). Natural language processing for information and project management. In Digital Transformation of the Design, Construction and Management Processes of the Built Environment, pp. 95-102. Springer. https://doi.org/10.1007/978-3-030-33570-0_9

Dicks, E. P., & Molenaar, K. R. (2022). Analysis of Washington State Department of Transportation Risks. Transportation Research Record, 03611981221109599. https://doi.org/10.1177/03611981221109599

Duijm, N.J., (2015). Recommendations on the use and design of risk matrices. Safety Science, 76, pp.21-31. https://doi.org/10.1016/j.ssci.2015.02.014

El-Sayegh, S.M. and Mansour, M.H., (2015). Risk assessment and allocation in highway construction projects in the UAE. Journal of Management in Engineering, 31(6), pp.04015004. https://doi.org/10.1061/(ASCE)ME.1943-5479.0000365

Erfani, A., & Cui, Q. (2021). Natural Language Processing Application in Construction Domain: An Integrative Review and Algorithms Comparison. Computing in Civil Engineering, 26-33. https://doi.org/10.1061/9780784483893.004

Erfani, A., & Cui, Q. (2022). Predictive risk modeling for major transportation projects using historical data. Automation in Construction, 139, 104301. https://doi.org/10.1016/j.autcon.2022.104301

Erfani, A., & Tavakolan, M., (2020). Risk Evaluation Model of Wind Energy Investment Projects Using Modified Fuzzy Group Decision-making and Monte Carlo Simulation. Arthaniti: Journal of Economic Theory and Practice, pp. 0976747920963222. https://doi.org/10.1177%2F0976747920963222

Erfani, A., Tavakolan, M., Mashhadi, A. H., & Mohammadi, P. (2021a). Heterogeneous or homogeneous? A modified decision-making approach in renewable energy investment projects. AIMS Energy, 9(3), 558-580. https://doi.org/10.3934/energy.2021027

Erfani, A., Cui, Q., & Cavanaugh, I. (2021b). An Empirical Analysis of Risk Similarity among Major Transportation Projects Using Natural Language Processing. Journal of Construction Engineering and Management, 147(12), 04021175. https://doi.org/10.1061/(ASCE)CO.1943-7862.0002206

Erfani, A., Zhang, K., & Cui, Q. (2021c). TAB bid irregularity: Data-driven model and its application. Journal of Management in Engineering, 37(5), 04021055. https://doi.org/10.1061/(ASCE)ME.1943-5479.0000958

Erfani, A., Hickey, P. J., & Cui, Q. (2023a). Likeability vs. Competence dilemma: A Text Mining approach using LinkedIn data. Journal of Management in Engineering.

Erfani, A., Ma, Z., Cui, Q., & Baecher, G. B. (2023b). Ex Post Project Risk Assessment: Method and Empirical Study. Journal of Construction Engineering and Management, 149(2), 04022174. https://doi.org/10.1061/JCEMD4.COENG-12588

Erfani, A., Cui, Q., Baecher, G., Kwak, Y. H. (2023c). Data-driven approach to risk identification in major transportation projects: A common risk breakdown structure, IEEE Transactions on Engineering Management.

Erfani, A., Villeda, V. H., & Cui, Q. (2022). Artificial Intelligence Application for Risk Template Generation in Major Transportation Projects. In Construction Research Congress 2022 (pp. 21-29). https://doi.org/10.1061/9780784483961.003

Ermentrout, G. B., & Edelstein-Keshet, L. (1993). Cellular automata approaches to biological modeling. Journal of theoretical Biology, 160(1), 97-133. https://doi.org/10.1006/jtbi.1993.1007

Fan, H., & Li, H. (2013). Retrieving similar cases for alternative dispute resolution in construction accidents using text mining techniques. Automation in construction, 34, 85-91. https://doi.org/10.1016/j.autcon.2012.10.014

Fang, W., Luo, H., Xu, S., Love, P. E., Lu, Z., & Ye, C. (2020). Automated text classification of near-misses from safety reports: An improved deep learning approach. Advanced Engineering Informatics, 44, 101060.

Federal Highway Administration (FHWA)., (2021). Major Projects. https://www.fhwa.dot.gov/majorprojects

Flood, M. D., & Goodenough, O. R. (2022). Contract as automaton: representing a simple financial agreement in computational form. Artificial Intelligence and Law, 30(3), 391-416. https://doi.org/10.1007/s10506-021-09300-9

Freire, J. G., & DaCamara, C. C. (2019). Using cellular automata to simulate wildfire propagation and to assist in fire management. Natural hazards and earth system sciences, 19(1), 169-179. https://doi.org/10.5194/nhess-19-169-2019

Ge, L., & Moh, T. S. (2017). Improving text classification with word embedding. In 2017 IEEE International Conference on Big Data (Big Data) (pp. 1796-1805). IEEE. https://doi.org/10.1109/BigData.2017.8258123

Gondia, A., Siam, A., El-Dakhakhni, W. and Nassar, A.H., (2020). Machine learning algorithms for construction projects delay risk prediction. Journal of Construction Engineering and Management, 146(1), pp.04019085. https://doi.org/10.1061/(ASCE)CO.1943-7862.0001736

Hassan, F. U., & Le, T. (2020). Automated Requirements Identification from Construction Contract Documents Using Natural Language Processing. Journal of Legal Affairs and Dispute Resolution in Engineering and Construction, 12(2), 04520009. https://doi.org/10.1061/(ASCE)LA.1943-4170.0000379

Hastak, M., & Shaked, A. (2000). ICRAM-1: Model for international construction risk assessment. Journal of management in engineering, 16(1), 59-69. https://doi.org/10.1061/(ASCE)0742-597X(2000)16:1(59)

Heravi, G., Taherkhani, A. H., Sobhkhiz, S., Mashhadi, A. H., & Zahiri-Hashemi, R. (2021). Integrating risk management's best practices to estimate deep excavation projects' time and cost. Built Environment Project and Asset Management. https://doi.org/10.1108/BEPAM-11-2020-0180

Hickey, P. J., Erfani, A., & Cui, Q. (2022). Use of LinkedIn Data and Machine Learning to Analyze Gender Differences in Construction Career Paths. *Journal of Management in Engineering*, *38*(6), https://doi.org/10.1061/JMENEA.MEENG-5213

Iqbal, S., Choudhry, R. M., Holschemacher, K., Ali, A., & Tamošaitienė, J. (2015). Risk management in construction projects. Technological and economic development of economy, 21(1), 65-78.

Islam, M. S., Nepal, M. P., & Skitmore, M. (2019). Modified fuzzy group decision-making approach to cost overrun risk assessment of power plant projects. Journal of

Construction Engineering and Management, 145(2), 04018126. https://doi.org/10.1061/(ASCE)CO.1943-7862.0001593

James, F. (2019). A Risk Management Framework and A Generalized Attack Automata for IoT based Smart Home Environment. In 2019 3rd Cyber Security in Networking Conference (CSNet) (pp. 86-90). IEEE.

Jung, W., & Han, S. H., (2017). Which risk management is most crucial for controlling project cost? Journal of Management in Engineering, 33(5), 04017029. https://doi.org/10.1061/(ASCE)ME.1943-5479.0000547

Kaliprasad, M. (2006). Proactive risk management. Cost Engineering, 48(12), 26.

Kasnesis, P., Heartfield, R., Liang, X., Toumanidis, L., Sakellari, G., Patrikakis, C., & Loukas, G. (2021). Transformer-based identification of stochastic information cascades in social networks using text and image similarity. Applied Soft Computing, 108, 107413.

Kenton, J. D. M. W. C., & Toutanova, L. K. (2019). BERT: Pre-training of Deep Bidirectional Transformers for Language Understanding. In Proceedings of NAACL-HLT (pp. 4171-4186).

Kim, T., & Chi, S. (2019). Accident case retrieval and analyses: using natural language processing in the construction industry. Journal of Construction Engineering and Management, 145(3). https://doi.org/10.1061/(ASCE)CO.1943-7862.0001625

Krippendorff, K. (2018). Content analysis: An introduction to its methodology. Sage publications.

Lauriola, I., Lavelli, A. and Aiolli, F., (2022). An introduction to deep learning in natural language processing: models, techniques, and tools. Neurocomputing, 470, pp.443-456. https://doi.org/10.1016/j.neucom.2021.05.103

Leva, M. C., Balfe, N., McAleer, B., & Rocke, M. (2017). Risk registers: Structuring data collection to develop risk intelligence. Safety science, 100, 143-156. https://doi.org/10.1016/j.ssci.2017.05.009

Li, L., Erfani, A., Wang, Y., & Cui, Q. (2021). Anatomy into the battle of supporting or opposing reopening amid the COVID-19 pandemic on Twitter: A temporal and spatial analysis. Plos one, 16(7), e0254359.

Linton, J., (2018). High-Speed Rail Cost Overrun Reporting Raises Questions of Media Bias. https://cal.streetsblog.org/2018/01/24/high-speed-rail-cost-overrun-reporting-raises-questions-of-media-bias/, Accessed September 20, 2022

Mackenzie, D. (1995). The automation of proof: A historical and sociological exploration. IEEE Annals of the History of Computing, 17(3), 7-29. https://doi.org/10.1109/85.397057

Mikolov, T., Sutskever, I., Chen, K., Corrado, G.S. and Dean, J., (2013). Distributed representations of words and phrases and their compositionality. Advances in Neural Information Processing Systems, 26.pp. 3111-3119.

Mohammadi, P., Ramezanianpour, A. M., & Erfani, A. (2022). Identifying and Prioritizing Criteria for Selecting Sustainable Façade Materials of High-Rise Buildings. Construction Research Congress 2022. https://doi.org/10.1061/9780784483978.060

Mohammadi, P., Rashidi, A., Malekzadeh, M., & Tiwari, S. (2023). Evaluating various machine learning algorithms for automated inspection of culverts. *Engineering Analysis with Boundary Elements*, *148*, 366-375.

Molenaar, K. R. (2010). Guidebook on risk analysis tools and management practices to control transportation project costs (Vol. 658). Transportation Research Board.

Molenaar, K. R., (2006). Guide to risk assessment and allocation for highway construction management. Federal Highway Administration.

Montana State Department of Transportation. (2016) Project Risk Management Guidelines: Managing project costs through identification and management of risks.

Montibeller, G. and Von Winterfeldt, D., (2015). Cognitive and motivational biases in decision and risk analysis. Risk Analysis, 35(7), pp.1230-1251. https://doi.org/10.1111/risa.12360

Monzer, N., Fayek, A. R., Lourenzutti, R., & Siraj, N. B. (2019). Aggregation-based framework for construction risk assessment with heterogeneous groups of experts. Journal of Construction Engineering and Management, 145(3), 04019003. https://doi.org/10.1061/(ASCE)CO.1943-7862.0001614

Moon, S., Chi, S., & Im, S. B. (2022). Automated detection of contractual risk clauses from construction specifications using bidirectional encoder representations from transformers (BERT). Automation in Construction, 142, 104465.

Morteza, A., Yahyaeian, A. A., Mirzaeibonehkhater, M., Sadeghi, S., Mohaimeni, A., & Taheri, S. (2023). Deep learning hyperparameter optimization: Application to electricity and heat demand prediction for buildings. Energy and Buildings, 289, 113036.

Nguyen, D. A., Garvin, M. J., & Gonzalez, E. E. (2018). Risk allocation in US public-private partnership highway project contracts. Journal of Construction Engineering and Management, 144(5). https://doi.org/10.1061/(ASCE)CO.1943-7862.0001465

O'Regan, G. (2021). Automata theory. In Guide to Discrete Mathematics (pp. 121-131). Springer, Cham. https://doi.org/10.1007/978-3-030-81588-2_7

Panahi, R., Louis, J., Aziere, N., Podder, A., & Swanson, C. (2022). Identifying Modular Construction Worker Tasks Using Computer Vision. In Computing in Civil Engineering 2021 (pp. 959-966).

Panahi, R.; Louis, J., Kivlin, J. (2023) Request for Information (RFI) Recommender System for Pre-Construction Design Review Application Using Natural Language Processing, Chat-GPT, and Computer Vision; In Computing in Civil Engineering 2023.

Partnerships, V. P. P. (2015). P3 Risk Management Guidelines. The Commonwealth.

Pavlak, A. (2004). Project troubleshooting: tiger teams for reactive risk management. Project Management Journal, 35(4), 5-14. https://doi.org/10.1177/875697280403500403

Pennington, J., Socher, R. and Manning, C.D., (2014). Glove: Global vectors for word representation. In Proceedings of the 2014 Conference on Empirical Methods in Natural Language Processing, pp. 1532-1543.

Rasool, M., Franck, T., Denys, B., & Halidou, N. (2012). Methodology and tools for risk evaluation in construction projects using Risk Breakdown Structure. European journal of environmental and civil engineering, 16(sup1), s78-s98. https://doi.org/10.1080/19648189.2012.681959

Reimers, N., & Gurevych, I. (2019). Sentence-BERT: Sentence Embeddings using Siamese BERT-Networks. In Proceedings of the 2019 Conference on Empirical Methods in Natural Language Processing and the 9th International Joint Conference on Natural Language Processing (EMNLP-IJCNLP) (pp. 3982-3992).

Rencher, A. C. (2005). A review of "Methods of Multivariate Analysis". https://doi.org/10.1080/07408170500232784

Richter, C. W., Sheblé, G. B., & Ashlock, D. (1999). Comprehensive bidding strategies with genetic programming/finite state automata. IEEE Transactions on Power systems, 14(4), 1207-1212. https://doi.org/10.1109/59.801874

Sanchez-Cazorla, A., Alfalla-Luque, R., & Irimia-Dieguez, A. I. (2016). Risk identification in megaprojects as a crucial phase of risk management: A literature review. Project Management Journal, 47(6), 75-93. https://doi.org/10.1177/875697281604700606

Sanni-Anibire, M.O., Zin, R.M. and Olatunji, S.O., (2020). Machine learning model for delay risk assessment in tall building projects. International Journal of Construction Management, pp.1-10. https://doi.org/10.1080/15623599.2020.1768326

Shahmirzadi, O., Lugowski, A., & Younge, K. (2019). Text similarity in vector space models: a comparative study. In 2019 18th IEEE International Conference on

Machine Learning And Applications (ICMLA) (pp. 659-666). IEEE. https://doi.org/10.1109/ICMLA.2019.00120

Shannon, C. E. (1953). Computers and automata. Proceedings of the IRE, 41(10), 1234-1241. https://doi.org/10.1109/JRPROC.1953.274273

Shimura, K., & Nishinari, K. (2014). Project Management and Critical Path Analysis: A Cellular Automaton Model. Journal of Cellular Automata, 9.

Sidorov, G. (2019). Vector Space Model for Texts and the tf-idf Measure. In Syntactic n-grams in Computational Linguistics (pp. 11-15). Springer, Cham. https://doi.org/10.1007/978-3-030-14771-6_3

Sigmund, Z., & Radujković, M. (2014). Risk breakdown structure for construction projects on existing buildings. Procedia-Social and Behavioral Sciences, 119, 894-901. https://doi.org/10.1016/j.sbspro.2014.03.100

Siraj, N. B., & Fayek, A. R., (2019). Risk identification and common risks in construction: Literature review and content analysis. Journal of Construction Engineering and Management, 145(9), 03119004. https://doi.org/10.1061/(ASCE)CO.1943-7862.0001685

Slowey, K., (2019). Delays bring Maryland Purple Line costs to $6 B. https://www.constructiondive.com/news/delays-bring-maryland-purple-line-costs-close-to-6b/551301/, Accessed September 20, 2022

Somi, S., Seresht, N.G. and Fayek, A.R., (2021). Developing a risk breakdown matrix for onshore wind farm projects using fuzzy case-based reasoning. Journal of Cleaner Production, 311, pp.127572. https://doi.org/10.1016/j.jclepro.2021.127572

Song, Z., Sun, F., Zhang, R., Du, Y., & Zhou, G. (2021). An Improved Cellular Automaton Traffic Model Based on STCA Model Considering Variable Direction Lanes in I-VICS. Sustainability, 13(24), 13626. https://doi.org/10.3390/su132413626

Taroun, A. (2014). Towards a better modelling and assessment of construction risk: Insights from a literature review. International journal of Project management, 32(1), 101-115. https://doi.org/10.1016/j.ijproman.2013.03.004

Tavakolan, M., & Etemadinia, H. (2017). Fuzzy weighted interpretive structural modeling: Improved method for identification of risk interactions in construction projects. Journal of Construction engineering and Management, 143(11), 04017084. https://doi.org/10.1061/(ASCE)CO.1943-7862.0001395

Tuohy, J., (2020). SR 37 project in Fishers still faces up to $42 million in cost overruns despite savings. https://www.indystar.com/story/news/local/hamilton-county/2020/11/17/fishers-state-road-37-project-still-faces-up-42-million-cost-overruns/6325749002/ , Accessed September 20, 2022

Vardi, M. Y. (1989). A note on the reduction of two-way automata to one-way automata. Information Processing Letters, 30(5), 261-264.

Wang, Y., Zhang, K., Liang, M., & Cui, Q. (2022). Identifying Contingency Liability from P3 Contracts Using Rule-Based NLP. In *Construction Research Congress 2022* (pp. 59-68).

Washington DOT, Strategic Analysis and Estimating Office. (2018) Project Risk Management Guide. Washington State Department of Transportation, Olympia.

Wen, Y., & Ray, A. (2012). Vector space formulation of probabilistic finite state automata. Journal of Computer and System Sciences, 78(4), 1127-1141. https://doi.org/10.1016/j.jcss.2012.02.001

Wicks, D. (2017). The coding manual for qualitative researchers. Qualitative research in organizations and management: an international journal, 12(2), 169-170. https://doi.org/10.1108/QROM-08-2016-1408

Zhang, F., (2019). A hybrid structured deep neural network with Word2Vec for construction accident causes classification. International Journal of Construction Management, pp.1-21. https://doi.org/10.1080/15623599.2019.1683692

Zhang, K., Erfani, A., Beydoun, O., & Cui, Q. (2022). Procurement Benchmarks for Major Transportation Projects. Transportation Research Record, 2676(11), 363-376. https://doi.org/10.1177/03611981221092722

Zhang, T., Kishore, V., Wu, F., Weinberger, K. Q., & Artzi, Y. (2019). BERTScore: Evaluating Text Generation with BERT. In International Conference on Learning Representations.

Zhong, B., Pan, X., Love, P.E., Ding, L. and Fang, W., (2020). Deep learning and network analysis: Classifying and visualizing accident narratives in construction. Automation in Construction, 113, pp.103089. https://doi.org/10.1016/j.autcon.2020.103089